\documentclass[iop]{emulateapj}
\usepackage{ulem}
\usepackage{longtable}
\usepackage{graphicx}
\usepackage{aas_macros}
\usepackage{epsfig}
\usepackage{apjfonts}

\shorttitle{Hyperaccreting Black Hole as GRB Central Engine II}
\shortauthors{Lei et al.}

\begin{document}
\title{Hyperaccreting Black Hole as Gamma-Ray Burst Central Engine. II. Temporal evolution of central engine parameters during Prompt and Afterglow Phases}

\author{Wei-Hua Lei$^{1,2}$, Bing Zhang$^{2}$, Xue-Feng Wu$^{3}$, and En-Wei Liang$^{4}$}
\affil{$^{1}$School of Physics, Huazhong University of Science and Technology, Wuhan 430074, China. Email: leiwh@hust.edu.cn \\
$^{2}$Department of Physics and Astronomy, University of Nevada Las Vegas, NV 89154, USA. Email: zhang@physics.unlv.edu  \\
$^{3}$Purple Mountain Observatory, Chinese Academy of Sciences, Nanjing 210008, China\\
$^{4}$Guangxi Key Laboratory for Relativistic Astrophysics, Department of Physics, Guangxi University, Nanning 530004, China \\
}

\begin{abstract}
A hyperaccreting stellar-mass black hole has been proposed as the candidate central engine of gamma-ray bursts (GRBs). The rich observations of GRBs by \textit{Fermi} and \textit{Swift} make it possible to constrain the central engine model by comparing the model predications against data. This paper is dedicated to studying the temporal evolution of central engine parameters for both prompt emission and afterglow phases. We consider two jet launching mechanisms, i.e., $\nu\bar{\nu}$ annihilations and the Blandford-Znajek (BZ) processe, and obtain analytical solutions to these two models. We then investigate the black hole central engine parameters, such as the jet power, the dimensionless entropy $\eta$, and the central engine parameter $\mu_0=\eta (1+\sigma_0)$ (where $\sigma_0$ is the initial magnetization of the engine) at the base of the jet. The black hole may be spun up by accretion, or spun down by the BZ process, leaving imprints in GRB lightcurves. Usually, a BZ jet is more powerful and is likely responsible for the late time central engine activities. However, an initially non-spinning black hole central engine may first launch a thermal ``fireball'' via neutrino annihilations, and then launch a Poynting-flux-dominated jet via the BZ process. Multiple flares, giant bumps, and plateaus in GRB afterglows can be well produced as the result of late time accretion onto the black hole.  
\end{abstract}


\section{Introduction}
The nature of the central engine of gamma-ray bursts (GRBs) remains a mystery. It is generally believed that long GRBs are connected with core-collapse supernovae (Woosley 1993; Paczy$\acute{n}$ski 1998;MacFadyen \& Woosley 1999), and short GRBs are likely related to mergers of two neutron stars or a neutron star and a black hole (Eichler et al. 1989; Paczy$\acute{n}$ski 1991; Fryer et al. 1999). These scenarios lead to the formation of a stellar mass black hole (BH) or a millisecond magnetar.

Two types of GRB central engine models have been discussed in the literature, i.e., the BH model and magnetar model. One popular model invokes a stellar-mass BH surrounded by a neutrino-cooling-dominated accretion flow (NDAF). Two mechanisms are considered to power the relativistic jet in a GRB for a BH central engine: the neutrino-antineutrino annihilation mechanism, which liberates the gravitational energy from the accretion disk (Popham et al. 1999, hereafter PWF99; Di Matteo et al. 2002, hereafter DPN02; Gu et al. 2006; Chen \& Beloborodov 2007; Janiuk et al. 2007; Lei et al. 2009; Liu et al. 2015); and the Blandford-Znajek (Blandford \& Znajek 1977, hereafter BZ) mechanism, which extracts the spin energy from the Kerr BH (Lee et al. 2000; Li 2000; Lei et al. 2013).

Thanks to \textit{Swift} and \textit{Fermi}, the observations have collected rich information on GRBs, which put further constraints on the GRB central engine models. For example, since a good fraction of GRBs are followed by X-ray flares (some have giant bumps and plateaus), the GRB central engine must be long-lived. In some GRBs (e.g., GRB 080916C), the broadband spectra show no evidence of quasi-thermal emission from a fireball photoshpere (Abdo et al. 2009), suggesting that at least for some GRBs, the central engine has to be strongly magnetized (Zhang \& Pe'er 2009). These observational constraints motivate us to systematically investigate the GRB BH central engine models. We planned to present our results in two papers. In Paper I (Lei et al. 2013, hereafter Paper I), we addressed the fundamental problem of baryon loading in GRB jets. We found that a magnetically dominated jet can be much cleaner and is more consistent with the requirement of large Lorentz factors in GRBs (Paper I). With the estimated Lorentz factor from the baryon loading rate, Yi et al. (2017) and Xie et al. (2017) found that some empirical correlations, such as jet power vs. Lorentz factor $\Gamma_0$ (Liang et al. 2010; Liang et al. 2015; Ghirlanda et al. 2012; L\"u et al. 2012) and minimum variability timescale (MTS, Wu et al. 2016) vs. the Lorentz factor $\Gamma_0$, favor the scenario in which the jet is driven by the BZ mechanism. A direct comparison between NDAF and BZ processes have been discussed, mostly considering the energy output only, in a number of works (PWF99; Kawanaka et al. 2013; Liu et al. 2015). However, a dedicated study on the evolution of central engine parameters, especially the baryon-loading-related dimensionless ``entropy'' $\eta$ (for the neutrino model), the magnetization parameter $\sigma_0$, and the central engine parameter $\mu_0=\eta (1+\sigma_0)$ (for BZ model), are still lacking. In the observational front, the temporal behavior of GRBs in the prompt emission and early afterglow phases may provide meaningful clues to the central engine models. It is therefore interesting to compare the predictions from the BH central engine models with the temporal behaviour of GRBs. This is the purpose of this Paper II. We continue to investigate the evolution of the BH central engine based on Paper I. 

This paper is organised as follows. In Section 2, we will study the two jet launching mechanisms within the context of Kerr metric in detail. We then apply our results to the prompt emission phase in Section 3 and the late central engine activity in Section 4. Finally, we summarize our results and discuss some related issues in Section 5.

\section{Black Hole Central Engine Model: Neutrino annihilation and Magnetic Powers}
For a spinning BH with hyper-accretion disk, energy can be extracted to power GRB by neutrino annihilations from the NDAF or by the BZ mechanism from the rotating BH. In this section, we will study these two mechanisms in detail. 

\subsection{Neutrino Model}
The neutrino model as the central engine of GRBs has been widely discussed (PWF99; Narayan, Piran \& Kumar 2001, hereafter NPK01; Kohri \& Mineshige 2002; DPN02; Chen \& Beloborodov 2007; Janiuk et al. 2004, 2007; Gu et al. 2006; Liu et al. 2007, 2015; Lei et al. 2009; Xie et al. 2016; for a review see Liu et al. 2017). The typical mass accretion rate in such a model is high ($0.01$ to $10 M_{\sun} s^{-1}$). Under such a condition, the gas photon opacity is also very high and radiation becomes trapped (Katz 1977; Begelman 1978; Abramowicz et al. 1988). However, neutrinos can still escape and tap the thermal energy of the disk produced by viscous dissipation before being advected into the BH. In this model, GRBs are powered by the energy liberated via the $\nu\bar{\nu}\rightarrow e^+e^-$ process in regions of low baryon density. 

DPN02 showed that the neutrino emission will be greatly suppressed by neutrino trapping for an accretion rate $\dot{M} \geq 1 M_{\sun} s^{-1}$. However, their results are based on a Newtonian disk model. Gu et al. (2006), Chen \& Beloborodov (2007) and Lei et al. (2010) argued that the general relativistic effects are also important. In this paper, we adopt a model of a steady-state disk around a Kerr BH, in which neutrino loss and transfer are taken into account.

The accretion rate likely varies at the central engine of a GRB. As a first step, we assume a constant mass accretion rate to get the general properties of an NDAF, and leave the study of the evolution of the disk in Sections 3 and 4.

Because the gas cools efficiently, we are entitled to discuss the NDAF model within the context of a thin disk (Sharkura \& Sunyaev 1973). The accuracy of the thin-disk approximation is not perfect at large radii, where the disk is thick. On the other hand, the details of the outer region have little effect on the solution for the neutrino-cooled disk (Chen \& Beloborodov 2007).

The basic equations of NDAF (equations for continuity, state, conservation of  angular momentum and energy balance) in the Kerr metric are given as follows (PWF99; DNP02; Reynoso, Romero \& Sampayo 2006; Lei et al. 2009),
\begin{equation}
\dot {M} = - 4\pi r v_r \rho H,
\label{eq:GRNDAF1}
\end{equation}

\begin{equation}
\dot{M} \sqrt {G M_\bullet r } \frac{D}{A} =  4\pi r^2H\alpha P\sqrt {\frac{A}{BC}},
\label{eq:GRNDAF2}
\end{equation}

\begin{eqnarray}
P = &  & \frac{11}{12}  aT^4 + \frac{\rho kT}{m_{\rm p} } \left(\frac{1 + 3X_{nuc} }{4}\right) +
\frac{2\pi hc}{3} \left(\frac{3}{8\pi m_{\rm p} } \right)^{4 / 3}  \nonumber \\ 
& & \left(\frac{\rho }{\mu _{\rm e} } \right)^{4 /3} + \frac{u_\nu }{3},
\label{eq:GRNDAF3}
\end{eqnarray}

\begin{equation}
Q^ + = Q^ - ,
\label{eq:GRNDAF4}
\end{equation}
where $H=\sqrt{Pr^3 B/(\rho G M_\bullet C)}$ is the half thickness of the disk, $v_r$ is the radial velocity of the gas, $\alpha$ is the viscosity parameter, $a$ is the radiation constant, $k$ is the gas Boltzmann constant, and $m_{\rm p}$ is the proton rest mass. $A, B, C, D$ and $f$ are the general relativistic correction factors for a thin accretion disk around a Kerr BH (Riffert {\&} Herold 1995).
\begin{equation}
\label{eqA}
A = 1 - \frac{2G M_\bullet}{c^2r} + (\frac{G M_\bullet a_\bullet }{c^2r})^2,
\end{equation}

\begin{equation}
\label{eqB}
B = 1 - \frac{3G M_\bullet}{c^2r} + 2a_\bullet (\frac{G M_\bullet}{c^2r})^{3 / 2},
\end{equation}

\begin{equation}
\label{eqC}
C = 1 - 4a_\bullet (\frac{G M_\bullet}{c^2r})^{3 / 2} + 3(\frac{G M_\bullet a_\bullet
}{c^2r})^2,
\end{equation}

\begin{equation}
\label{eqD}
D = B f,
\end{equation}
where the BH spin parameter $a_\bullet=J_\bullet c/G M_\bullet^2$, and $M_{\bullet}$ and $J_\bullet$ are the BH mass and angular momentum, respectively. The expression for $f$ is given by Page \& Thorne (1974) (in their Equation (15n)) as,
\begin{eqnarray}
f & = & \frac{\chi^2}{(\chi^3-3\chi+2a_\bullet)} [\chi-\chi_{\rm ms} -\frac{3}{2}a_\bullet \ln(\frac{\chi}{\chi_{\rm ms}})- \nonumber \\
&  & \frac{3(\chi_1-a_\bullet)^2}{\chi_1(\chi_1-\chi_2)(\chi_1-\chi_3)} \ln(\frac{\chi-\chi_1}{\chi_{\rm ms}-\chi_1}) -  \nonumber \\
&  & \frac{3(\chi_2-a_\bullet)^2}{\chi_2(\chi_2-\chi_1)(\chi_2-\chi_3)} \ln(\frac{\chi-\chi_2}{\chi_{\rm ms}-  \chi_2})- \nonumber \\
&  &  \frac{3(\chi_3-a_\bullet)^2}{\chi_3(\chi_3- \chi_1)(\chi_3-\chi_2)} \ln(\frac{\chi-\chi_3}{\chi_{\rm ms}-\chi_3}) ],
\end{eqnarray}
where $\chi=(r/r_{\rm g})^{1/2}$, $\chi_{\rm ms}=(r_{\rm ms}/r_{\rm g})^{1/2}$, and $r_{\rm g}=G M_\bullet/c^2$. The radius of the marginally stable orbit is (Bardeen et al. 1972)
\begin{equation}
r_{\rm ms}=r_{\rm g} [3+Z_2 - sgn(a_\bullet)[(3-Z_1)(3+Z_1+2Z_2)]^{1/2}], 
\end{equation}
for $0\leq a_{\bullet} \leq 1$, where $Z_1 \equiv 1+(1-a_\bullet^2)^{1/3} [(1+a_\bullet)^{1/3}+(1-a_\bullet)^{1/3}], \ \ Z_2\equiv (3a_\bullet^2+Z_1^2)^{1/2}$, and $\chi_1= 2 \cos(\frac{1}{3} \cos^{-1}a_\bullet - \pi/3)$, $\chi_2= 2 \cos(\frac{1}{3} \cos^{-1}a_\bullet + \pi/3)$, $\chi_3=- 2 \cos(\frac{1}{3} \cos^{-1}a_\bullet )$ are the three roots of $\chi^3 - 3\chi+ 2 a_\bullet = 0$. It is easy to check that $f(r=r_{\rm ms})=0$ and $f(r \gg r_{\rm ms})\sim 1-\sqrt{r_{\rm ms}/r}$. 

In Equation (\ref{eq:GRNDAF3}), the total pressure consists of four terms, radiation pressure,
gas pressure, degeneracy pressure, and neutrino pressure. The factor $11/12$ in the term of radiation pressure includes the contribution of relativistic electron-positron pairs. In the degeneracy pressure
term, $\mu_e$ is the mass per electron, which is taken as 2 in agreement with NPK and PWF. $u_\nu $ is the neutrino energy density defined as (Popham {\&} Narayan 1995)
\begin{equation}
u_\nu = (7 / 8)aT^4\sum {\frac{\tau _{\nu _i } / 2 + 1 / \sqrt 3 }{\tau
_{\nu _i } / 2 + 1 / \sqrt 3 + 1 / (3\tau _{a,\nu _i } )}},
\label{eq:uv}
\end{equation}
where $\tau _{\nu _i } = \tau _{a,\nu _i } + \tau _{s,\nu _i } $
is the sum of the absorptive and scattering optical depths calculated for each
neutrino flavor $(\nu _e ,\nu _\mu ,\nu _\tau )$. The absorptive optical
depths for the three neutrino flavors are (Kohri et al. 2005)

\begin{equation}
\tau _{a,\nu _e } \simeq 2.5\times 10^{ - 7}T_{11}^5 H + 4.5\times 10^{ -
7}T_{11}^2 X_{\rm nuc} \rho _{10} H,
\label{eq:tauae}
\end{equation}

\begin{equation}
\tau _{a,\nu _\mu } = \tau _{a,\nu _\tau } \simeq 2.5\times 10^{ -
7}T_{11}^5 H,
\label{eq:tauvv}
\end{equation}
\noindent where $T_{11}=T/10^{11} \rm K$, $\rho_{10} = \rho/10^{10} \rm g \ cm^{-3}$. $X_{\rm nuc} \simeq 34.8\rho _{10}^{ - 3 / 4} T_{11}^{9 / 8} \exp ( - 0.61 /T_{11} )$ is the mass fraction of free nucleons (PWF99; DPN02).

The total scattering optical depth is given by (DPN02)
\begin{equation}
\tau _{s,\nu _i } \simeq 2.7\times 10^{ - 7}T_{11}^2 \rho _{10} H.
\label{eq:taus}
\end{equation}

In Equation (\ref{eq:GRNDAF4}), $Q^ + = Q_{\rm vis} $ represents viscous
dissipation, and $Q^ - = Q_\nu + Q_{\rm photo} + Q_{\rm adv} $ is the total
cooling rate due to neutrino losses $Q_\nu $, photodisintegration
$Q_{\rm photo} $ and advection $Q_{\rm adv} $. We employ a bridging formula
for calculating $Q_\nu $, which is valid in both the optically thin
and thick cases. The expressions for $Q_\nu $, $Q_{\rm photo} $ and
$Q_{\rm adv} $ are (DPN02)

\begin{equation}
Q_\nu = \sum {\frac{(7 / 8\sigma T^4)}{(3 / 4)(\tau _{\nu _i } / 2 + 1 /
\sqrt 3 + 1 / (3\tau _{a,\nu _i } ))}} ,
\label{eq:Qv}
\end{equation}

\begin{equation}
Q_{\rm photo} = 10^{29}\rho _{10} v_r \frac{dX_{\rm nuc} }{dr}H\mbox{ } {\rm erg \cdot
cm^{-2}s^{-1} },
\label{eq:Qphoto}
\end{equation}

\begin{equation}
Q_{\rm adv} \simeq v_r \frac{H}{r}(\frac{11}{3}aT^4 + \frac{3}{2}\frac{\rho
kT}{m_{\rm p} }\frac{1 + X_{\rm nuc} }{4} + \frac{4u_\nu }{3}),
\label{eq:Qadv}
\end{equation}

The heating rate $Q_{\rm vis} $ is expressed as
\begin{equation}
Q_{\rm vis} = \frac{3G M_\bullet \dot{M} }{8\pi r^3} f.
\label{eq:Qvis} 
\end{equation}

We solve numerically Equations (\ref{eq:GRNDAF1})-(\ref{eq:Qvis}) to find the disk temperature $T$ and density $\rho $ versus the disk radius given
$a_\bullet $, $m$ and $\dot {m}$ (where $m_\bullet=M_\bullet/M_{\sun}$, and  $\dot {m}=\dot{M}/M_{\sun} s^{-1}$). We take $X_{\rm nuc} = 1$ for fully
photodisintegrated nuclei, which is appropriate in the inner disk. Furthermore, $\alpha = 0.1$ is adopted.

In the calculations, we ignore the cooling rate arising from
photodisintegration, because it is much less than the neutrino cooling rate
in the inner disk (Janiuk et al. 2004). We also approximately take the free nucleon fraction $X_{\rm nuc} \simeq 1$. For the disks formed by the collapses of massive stars, the photodisintegration process that breaks down $\alpha$-particles into neutrons and protons is important in the disk region at very large radii. However, the effect of photodisintegration becomes less significant for the region at small radii, which contains fewer $\alpha$-particles. See Kohri et al. (2005), Chen \& Beloborodov (2007) and Liu et al. (2007) for details, which showed that photodisintegration is not important for $r \leq 10^2 r_g$. On the other hand, for disks formed by the mergers of compact star binaries, we reasonably take all the nucleons to be free ($X_{\rm nuc} \simeq 1$) and neglect the photodisintegration process, since we mainly focus on the inner region of the disk.

The neutrino power from the accretion flow is given by,
\begin{equation}
\label{eq26}
\dot{E}_\nu = 4\pi \int_{r_{\rm ms} }^{r_{\rm out} } {Q_\nu rdr} .
\end{equation}

We are interested primarily in the properties of the inner accretion
flow, where neutrino processes are important. As argued by PWF99,
NPK01 and DPN02, for $r > 100r_g $, the neutrino cooling is not important and photons are
completely trapped. The flows are fully advection-dominated at that region. 
We therefore concentrate on discussing in the region from $r_{\rm ms} $ to $r_{\rm max } = 100 r_{\rm g} $.

In order to get the neutrino annihilation power, we model the disk as a grid of cells in the equatorial plane.
A cell $k$ has its neutrino mean energy
$\varepsilon _{\nu _i }^k $ and luminosity $l_{\nu _i }^k $, and the
height above (or below) the disk is $d_k $. The angle at which
neutrinos from cell $k$ encounter anti-neutrinos from another cell
$k'$ at that point is denoted as $\theta _{k{k}'} $. Then the
neutrino annihilation power at that point is given by the
summation over all pairs of cells (PWF99; Rosswog et al. 2003),

\begin{eqnarray}
\dot{E}_{\nu \bar{\nu}}  = & &  A_1 \sum_k \frac{l^k_{\nu_i}}{d_k^2} \sum_{k'}
\frac{l^k_{\nu_i}}{d_{k'}^2}
(\epsilon^k_{\nu_i}+\epsilon^{k'}_{\bar{\nu}_i})(1-cos\theta_{kk'})^2
 +   \nonumber \\
& & A_2 \sum_k \frac{l^k_{\nu_i}}{d_k^2} \sum_{k'} \frac{l^k_{\nu_i}}{d_{k'}^2} \frac{\epsilon^k_{\nu_i}+\epsilon^{k'}_{\bar{\nu}_i}}{\epsilon^k_{\nu_i} \epsilon^{k'}_{\bar{\nu}_i}}(1-cos\theta_{kk'})
\end{eqnarray}
where $A_1 \approx 1.7\times 10^{ - 44} \ {\rm cm \cdot erg^{-2} \cdot s^{-1} }$
and $A_2 \approx 1.6\times 10^{ - 56} \ {\rm cm \cdot erg^{ - 2}s^{-1} }$. 

\begin{figure*}[ht]
\center
\includegraphics[width=8cm,angle=0]{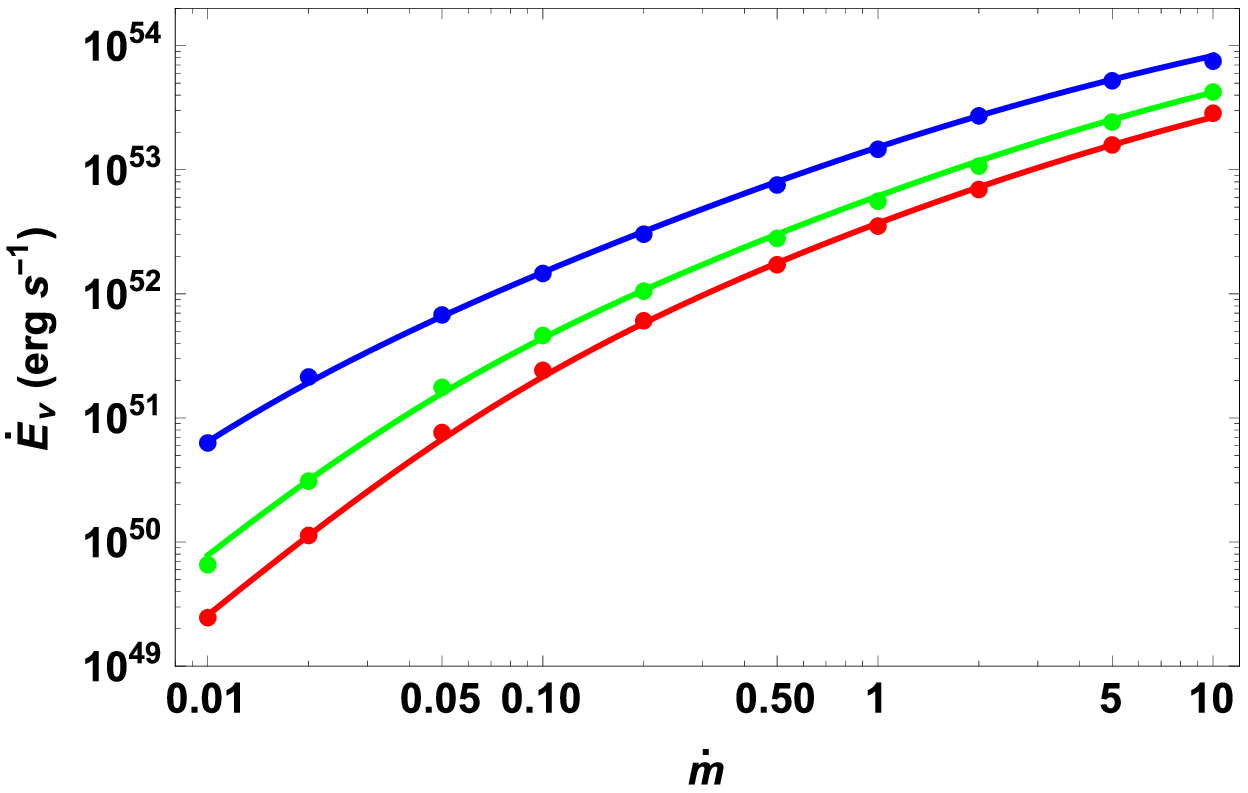}
\includegraphics[width=8cm,angle=0]{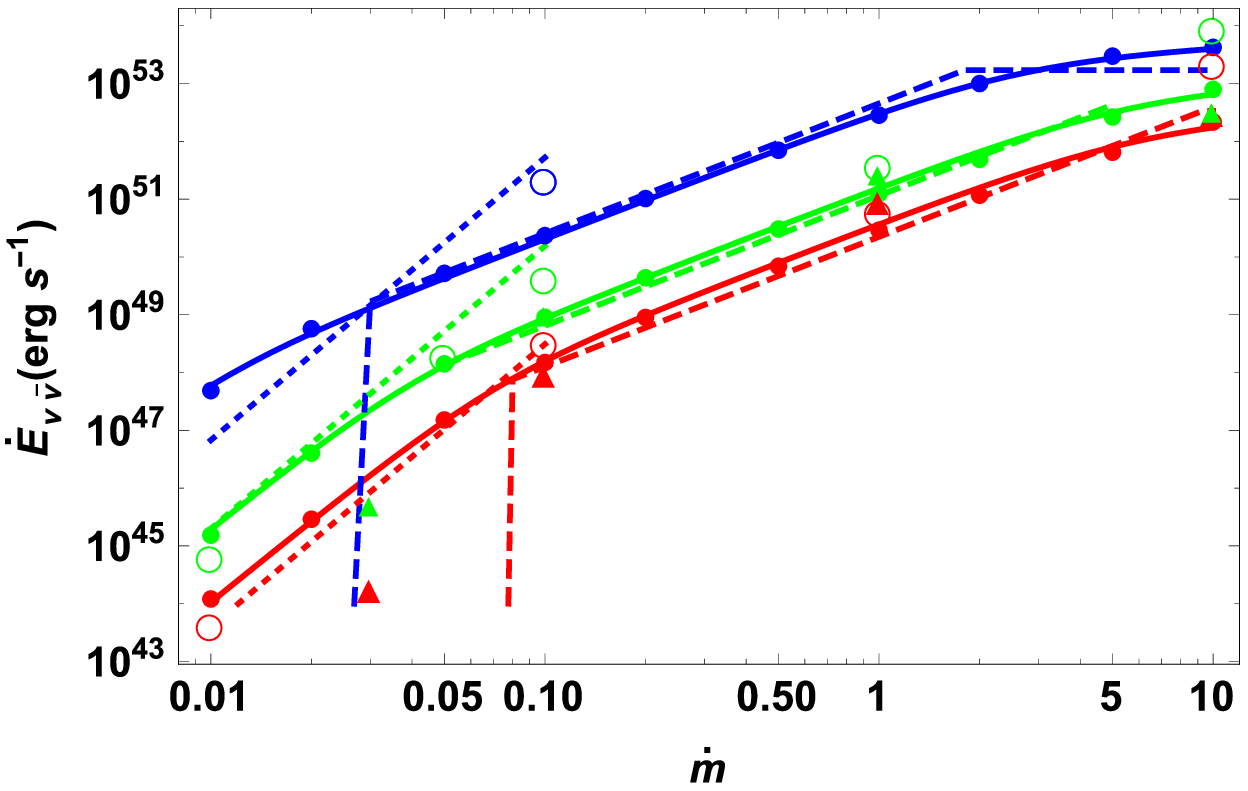}
\caption{The neutrino power $\dot{E}_{\nu}$ (left) and neutrino annihilation power $\dot{E}_{\nu\bar{\nu}}$ as a function of accretion rate for three different values of BH spin, $a_\bullet=0$ (red), $0.5$ (green ) and $0.95$(blue). Our analytical (see Equations. (\ref{eq_Ev})-(\ref{eq_Evv})) and numerical solutions are plotted with solids lines and points, respectively. For $\dot{E}_{\nu\bar{\nu}}$, we also show other analytical results from Zalamea \& Beloborodov (2011, dashed lines) and Fan et. (2005, dotted lines), and other numerical solutions from PWF99 (open circle) and Xue et al. (2013, filled triangle). In all the calculations, we adopt a BH with mass $m=3$. It is found that our analytical results agree well with the numerical solutions in all accretion rate regimes.}
\label{fig_dotEvv}
\end{figure*}

The total neutrino annihilation luminosity is obtained by integrating over
the whole space outside the BH and the disk. As a typical case, we show the results of neutrino power $\dot{E}_{\nu}$ (left panel) and neutrino annihilation power $\dot{E}_{\nu\bar{\nu}}$ (right panel) for a BH with mass $m_\bullet=3$ and with different accretion rate and BH spin in Figure \ref{fig_dotEvv} (points in the figure), in which $\alpha = 0.1 $ is adopted. For comparison, we also show the results from previous works, such as PWF99 (open symbols), and Xue et al. (2013, filled symbols). Inspecting Figure \ref{fig_dotEvv}, one finds that the results by PWF99 overestimate the neutrino annihilation power in the high accretion rate region. A reasonable understanding for this disagreement is the lack of neutrino trapping in PWF99 solutions. 

Generally, our resulting curves (thick dotted lines in Figure \ref{fig_dotEvv}) exhibit broken a power law shape with two breaks. The first break marks the transition of the inner disk from neutrino-dominated to advection-dominated. Following Chen \& Beloborodov (2007), we take the accretion rate at this break as $\dot{m}_{\rm ign}$, i.e., the disc temperature is not high enough to ignite the neutrino emitting reactions if $\dot{m} <\dot{m}_{\rm ign}$. The second break is due to the neutrino trapping effects (see DPN02 for details), and the corresponding accretion rate is denoted by $\dot{m}_{\rm trap}$. If $\dot{m} > \dot{m}_{\rm trap}$, the emitted neutrinos become trapped in the disc and advected into the BH. Therefore, for convenience, we summarize our numerical results with smooth power law fits with two breaks (shown with solid lines in Figure \ref{fig_dotEvv}), i.e., 
\begin{eqnarray}
 \dot{E}_{\nu}   \simeq  && \dot{E}_{\nu, \rm ign} \left[\left(\frac{\dot{m}}{\dot{m}_{\rm ign}} \right)^{-\alpha_\nu } + \left(\frac{\dot{m}}{\dot{m}_{\rm ign} }\right)^{-\beta_\nu } \right]^{-1}   \nonumber \\  
&& \times  \left[1+\left(\frac{\dot{m}}{\dot{m}_{\rm trap} } \right)^{\beta_\nu - \gamma_\nu } \right]^{-1} ,
\label{eq_Ev}
\end{eqnarray}

\begin{eqnarray}
\dot{E}_{\nu \bar{\nu}}  \simeq  && \dot{E}_{\nu \bar{\nu}, \rm ign} \left[ \left(\frac{\dot{m}}{\dot{m}_{\rm ign}} \right)^{-\alpha_{\nu \bar{\nu}} } + \left(\frac{\dot{m}}{\dot{m}_{\rm ign} } \right)^{-\beta_{\nu \bar{\nu}} } \right]^{-1} \nonumber \\
&& \times   \left[1+(\frac{\dot{m}}{\dot{m}_{\rm trap} })^{\beta_{\nu \bar{\nu}} - \gamma_{\nu \bar{\nu}}  }\right]^{-1} ,
\label{eq_Evv}
\end{eqnarray}
where,
\begin{eqnarray}
&&\left\lbrace
\begin{tabular}{l}
$\dot{E}_{\nu, \rm ign}=10^{(51.4-0.3 a_\bullet^2)}  \left(\frac{m_\bullet}{3} \right)^{\log(\dot{m}/\dot{m}_{\rm ign}) -1.5} {\rm erg \ s^{-1}},  $  \\
$\alpha_\nu = 2.3, \ \beta_\nu =  1.12, \ \gamma_\nu = 0.4, $
\end{tabular} 
\right.\nonumber \\
&&\left\lbrace
\begin{tabular}{l}
$\dot{E}_{\nu \bar{\nu}, \rm ign}=10^{(48.0+0.15 a_\bullet)}  \left(\frac{m_\bullet}{3} \right)^{\log(\dot{m}/\dot{m}_{\rm ign}) -3.3} {\rm erg \ s^{-1}},  $ \\
$\alpha_{\nu \bar{\nu}} = 4.7, \ \beta_{\nu \bar{\nu}} =  2.23, \  \gamma_{\nu \bar{\nu}} =0.3, $
\end{tabular} 
\right.\nonumber \\
&& \dot{m}_{\rm ign} = 0.07-0.063 a_\bullet,  \ \dot{m}_{\rm trap} = 6.0-4.0 a_\bullet^3,
\end{eqnarray}
where $\dot{m}_{\rm ign}$ and $\dot{m}_{\rm trap}$ are the igniting and trapping accretion rates, respectively. For  $m_\bullet=3$ and $\alpha=0.1$, $\dot{m}_{\rm ign}=0.07$ and $\dot{m}_{\rm trap}=6.0$ for $a_\bullet=0$, and $\dot{m}_{\rm ign}=0.01$ and $\dot{m}_{\rm trap}=2.6$ for $a_\bullet=0.95$. Similar results are obtained by Kohri et al. (2005) and Chen \& Beloborodov (2007).\footnote{In Chen \& Beloborodov (2007), the characteristic accretion rates $\dot{m}_{\rm ign}$ and $\dot{m}_{\rm trap}$ are well approximated by the following formulae, $\dot{m}_{\rm ign}=K_{\rm ign} \alpha_{-1}^{5/3}$, and $\dot{m}_{\rm trap}=K_{\rm trap} \alpha_{-1}^{1/3} $. For $a_\bullet=0$, one has $K_{\rm ign}=0.071$ and $K_{\rm trap}=9.3$, whereas for $a_\bullet=0.95$, one has $K_{\rm ign}=0.021$ and $K_{\rm trap}=1.8$. } 

With the second terms in $\dot{E}_{\nu, \rm ign}$ and $\dot{E}_{\nu \bar{\nu}, \rm ign}$, our analytical solutions can also apply to an NDAF with the BH mass in the range from $m_\bullet=3$ to $10$. To illustrate the accuracy of these power law fits we compare our fits (solid lines) with the numerical solutions and with the analytic formula obtained by several other authors, such as Fan et al. (2005, thin dotted lines) \footnote{
Fan et al.(2005) found that the $\nu \bar{\nu}$ power for $0.01<\dot{m}<0.1$ can be well fitted  with (see the thin dotted lines in the right panel of Figure \ref{fig_dotEvv}), $ \dot{E}_{\nu \bar{\nu}}  \simeq  10^{43.6+4.3 a_\bullet} (\frac{\dot{m}}{0.01})^{4.89} {\rm erg \ s^{-1} } $.
}
and Zalamea \& Beloborodov (2011, thin dashed lines).
\footnote{ Zalamea \& Beloborodov (2011) also obtained a simple formula for $\dot{E}_{\nu\bar{\nu}}$ as: 
0 for $\dot{m}< \dot{m}_{\rm ign}$; 
$ 1.1\times 10^{52} (\frac{r}{r_{\rm ms}})^{-4.8} (\frac{m_\bullet}{3})^{-3/2} \dot{m}^{9/4}  $ for $ \dot{m}_{\rm ign}< \dot{m}<\dot{m}_{\rm trap}$; 
$ 1.1\times 10^{52} (\frac{r}{r_{\rm ms}})^{-4.8} (\frac{m_\bullet}{3})^{-3/2} \dot{m}_{\rm trap}^{9/4} $ for $ \dot{m}>\dot{m}_{\rm trap}$.  
For $\alpha=0.1$, one has $\dot{m}_{\rm ign} = 0.071$ and $\dot{m}_{\rm trap}=9.3$ for $a_\bullet= 0$, and $\dot{m}_{\rm ign} = 0.021$ and $\dot{m}_{\rm trap}=1.8$ for $a_\bullet= 0.95$.
} in Figure \ref{fig_dotEvv}.

From Figure \ref{fig_dotEvv}, we find that our analytical solution (Equation (\ref{eq_Evv})) agrees quite well with the analytical solution by Zalamea \& Beloborodov (2011) for $\dot{m}>\dot{m}_{\rm ign}$, and the numerical solution by PWF99 (or the analytical one by Fan et al. 2005) for small BH spin and low accretion rates. Zalamea \& Beloborodov (2011) did not treat the NDAF with $\dot{m}<\dot{m}_{\rm ign}$, and roughly setted $\dot{E}_{\nu\bar{\nu}}$ as constant for $\dot{m}>\dot{m}_{\rm trap}$. Fan et al. (2005) only fitted for $0.01<\dot{m}<0.1$. Our analytical solutions, however, cover all the three regions (the whole range of accretion rate) rather smoothly. Therefore, for convenience, we will adopt our analytical solutions (i.e., Equations (\ref{eq_Ev}) and (\ref{eq_Evv})) directly in the following calculations. 

The baryon loading of jet is the fundamental problem in GRBs. In Paper I (Lei et al. 2013), we obtained the baryon loading rate for the jet driven by neutrino-annihilation,
\begin{eqnarray}
\dot{M}_{\rm j, \nu\bar{\nu}}  & \simeq & 7.0 \times 10^{-7} A^{0.85} B^{-1.35} C^{0.22} \theta_{\rm j,-1}^2  \alpha_{-1}^{0.57} \epsilon_{-1}^{1.7} \nonumber \\
& & \left( \frac{R_{\rm ms} }{2} \right) ^{0.32} \dot{m}_{-1}^{1.7} \left(\frac{m_\bullet}{3}\right)^{-0.9} \left(\frac{\xi}{2} \right)^{0.32} \ M_{\sun} {\rm s}^{-1}. \nonumber \\
\end{eqnarray}
for $\dot{m}>\dot{m}_{\rm ign}$, where $\theta_{\rm j}$ is the jet half opening angle, $\xi \equiv r/r_{\rm ms}$ is the disk radius in uints of $r_{\rm ms}$, $\epsilon$ is the neutrino emission efficiency, i.e., $\epsilon = \dot{E}_{\nu} /\dot{M}c^2$. For $\dot{m} <\dot{m}_{\rm ign}$, the neutrino cooling becomes unimportant. The dependence of $\dot{M}_{\rm j, \nu\bar{\nu}} $ on the accretion rate $\dot{m}$ might be replaced with $\dot{M}_{\rm j, \nu\bar{\nu}} \propto \dot{m}^{3.8}$.

We can thus define an important quantity in GRB central engine, the dimensionless ``entropy'' parameter $\eta$ as
\begin{equation}
\eta \equiv \frac{\dot{E}_{\rm m}}{ \dot{M}_{\rm j, \nu\bar{\nu}} c^2 }.
\label{eta}
\end{equation}
where $\dot{E}_{\rm m} = \dot{E}_{\nu \bar{\nu}}+ \dot{M}_{\rm j, \nu\bar{\nu}} c^2$ is the total matter energy outflow luminosity. 

This $\eta$ parameter describes the maximum available Lorentz factor in neutrino annihilation model (supposing that the neutrino annihilation energy is totally converted into kinetic energy of baryons after acceleration), i.e., $\Gamma_{\rm max} \simeq \eta$. 

To evolve these central engine parameters (such as $\dot{E}_{\nu \bar{\nu}}$ and $\eta$) with time, we need to consider the evolution of BH, since most of these parameters have significant dependences on the BH spin. During the hyper-accreting process, the equations for BH evolution are, 
\begin{equation}
\frac{dM_\bullet c^2}{dt} = \dot{M} E_{\rm ms},
\label{dMvv}
\end{equation}

\begin{equation}
\frac{dJ_\bullet}{dt} = \dot{M} L_{\rm ms},
\label{dJvv}
\end{equation}
where $E_{\rm ms}$ and $L_{\rm ms}$ are the specific energy and the specific momentum corresponding to the inner most radius $r_{\rm ms}$ of the disk, which are defined in Novikov \& Thorne (1973) as
$E_{\rm ms} = (4\sqrt{ R_{\rm ms} }-3a_{\bullet}) /(\sqrt{3} R_{\rm ms})$, $L_{\rm ms} = (G M_\bullet/c) (2 (3 \sqrt{R_{\rm ms}} -2 a_\bullet) )/(\sqrt{3} \sqrt{R_{\rm ms}} )$, where $R_{\rm ms} = r_{\rm ms}/r_{\rm g} $.

As $a_\bullet = J_\bullet c/(G M_\bullet^2)$, by incorporating the above two equations, we find that the BH will be spun up by the accretion with a rate as
\begin{equation}
\frac{da_\bullet}{dt} =  \dot{M} L_{\rm ms} c/(G M_\bullet^2) -2 a_\bullet \dot{M}  E_{\rm ms} /(M_\bullet c^2)
\end{equation}

The duration of the burst, in such a model, is determined by the viscous timescale of the accreting gas. In most accretion flows, the viscous time is significantly longer than the dynamical time, so the accretion model naturally explains the large difference between the duration of bursts and their minimum variability timescales.

Another topic of NDAF is about its stability, since it will shape the GRB lightcurve. The stability properties of NDAFs were first discussed by NPK01. They found that their NDAF is unstable only if it is optically thin and radiation pressure dominated, which could conceivably play a role in determining the temporal behavior of some bursts. In other cases, their NDAF solution is viscously, thermally and gravitationally stable. After considering neutrino trapping, DPN02 found that NDAFs are viscously and thermally stable, but are only gravitationally unstable for an extremely large accretion rate like $\dot{m} \sim 10$ and for $r \geq 50$. By including microphysics and photodisintegration, Janiuk et al. (2007) suggested that for sufficiently large accretion rates ($\dot{m} \geq 10$), the inner regions of the disk become opaque and develop a viscous and thermal instability. However, these models did not consider the effect of magnetic fields. Lei et al. (2009) pointed out that an NDAF torqued by magnetic coupling is viscously and thermally unstable for $\dot{m} \geq 0.086$. Janiuk and Yuan (2010) extended their work by introducing the BH spin and magnetic field. It is shown that the instability can occur when $\dot{m} \geq 0.5$ for a fastly spinning BH. Recently, Xie et al. (2016) suggested that the inner-boundary torque should be taken into account for NDAFs, and obtained an unstable solution as a possible interpretation for the variability of GRB prompt emission and X-ray flares. Shibata et al. (2007), on the other hand, performed an axisymmetric general relativity magnetohydrodynamic (GRMHD) simulation for neutrino-cooled accretion tori around a rotating BH. Their results suggest that the angular momentum transport and the consequent shock heating caused by magnetic stress will induce a time-varying neutrino power, which is favorable for explaining the variability of GRB lightcurves.

\subsection{Magnetic Model}
Blandford \& Znajek (1977) proposed that the rotating energy and the angular momentum of a BH can be extracted by a surrounding magnetic field, and this energy mechanism has been referred to as the BZ process. If the magnetic field of BH is strong enough ($\sim 10^{15} {\rm G} $), the rotational energy extracted by this process can power GRBs (Paczy$\acute{n}$ski 1998; M\'{e}sz\'{a}ros \& Rees 1997; Paper I; Tchekhovskoy \& Giannios 2015). On the other hand, researches showed that the magnetic fields can be magnified up to $10^{15} \sim 10^{16} {\rm G}$ by virtue of MRI or dynamo process (Pudritz \& Fahlman 1982 and references therein) in hyperaccretion disks. 

The rotational energy of a BH with angular momentum $J_\bullet$ is a fraction of the BH mass $M_\bullet$,
\begin{equation}
E_{\rm rot}= 1.8 \times 10^{54} f_{\rm rot}(a_\bullet) \frac{M_\bullet}{M_\sun} {\rm erg},
\end{equation}

\begin{equation}
f_{\rm rot}(a_\bullet)=1-\sqrt{(1+\sqrt{1-a_\bullet^2})/2 },
\end{equation}
For a maximally rotating BH ($a_\bullet=1$), $f_{\rm rot}(1)=0.29$. 

The BZ jet power from a BH with mass $M_{\bullet}$ and angular momentum $J_\bullet$ is (Lee et al. 2000; Li 2000; Wang et al. 2002; McKinney 2005; Lei et al. 2005; Lei \& Zhang 2011; Lei et al. 2013) 
\begin{eqnarray}
\dot{E}_{\rm B}=1.7 \times 10^{50}  a_{\bullet}^2 m_{\bullet}^2
B_{\bullet,15}^2 F(a_{\bullet}) \ {\rm erg \ s^{-1}} \\ \nonumber
\simeq 1.1 \times 10^{50}  a_{\bullet}^2 m_{\bullet}^2
B_{\bullet,15}^2 \ {\rm erg \ s^{-1}},
\label{eq_Lmag}
\end{eqnarray}
where $B_{\bullet,15}=B_{\bullet}/10^{15} {\rm G}$ and $F(a_\bullet)=[(1+q^2)/q^2][(q+1/q) \arctan q-1]$. Here $q= a_{\bullet} /(1+\sqrt{1-a^2_{\bullet}})$, and $2/3\leq F(a_{\bullet}) \leq \pi-2$ for 
$0\leq a_{\bullet} \leq 1$. It apparently depends on $M_{\bullet}$, $B_{\bullet}$,
and $a_{\bullet}$. A strong magnetic field of the order $\sim 10^{15} \rm G$ is required to produce the high luminosity of a GRB. The accumulation of magnetic flux by an accretion flow may account for such a high magnetic field strength (e.g., Tchekhovskoy et al. 2011).

\begin{figure}[ht]
\center
\includegraphics[width=8cm,angle=0]{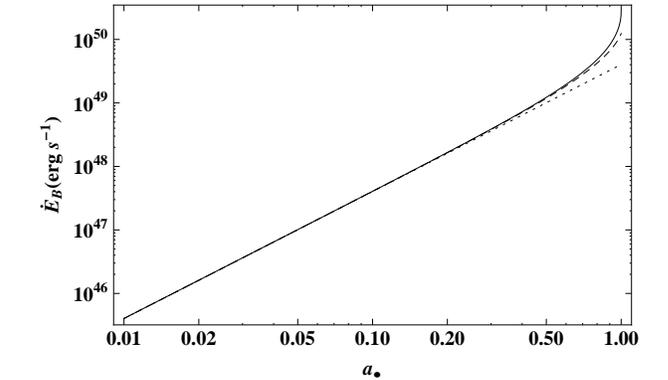}
\caption{The magnetic power $\dot{E}_{\rm B} $ as a function of BH spin $a_\bullet$. Solid line is the result with Equation (\ref{eq_Lmag}). We also plot the result from Tchekhovskoy, Narayan \& McKinney (2011) and BZ77 with dashed and dotted lines, respectively. In our calculations, we adopt the magnetic flux $\Phi_{\rm BH}=10^{27} \rm  G \ cm^2$, BH mass $m_\bullet=3$, and $\kappa=0.053$. }
\label{fig:dotEBa}
\end{figure}

The dependence of $\dot{E}_{\rm B}$ on BH spin is shown in Figure \ref{fig:dotEBa}. For comparison, we also plot the expressions given by BZ77 (derived in the limit $a_\bullet \ll 1$, but widely used, e.g., Thorne et al. 1986, PWF99)\footnote{BZ77 showed that magnetic power of a force-free jet from a slowly spinning BH ($a_\bullet \ll 1$) is $\dot{E}_{\rm B} = \frac{\kappa c}{4\pi} \Phi_{\rm BH}^2 \frac{a_\bullet^2}{16 r_{\rm g}^2}$, where $\kappa$ weakly depends on the field geometry (it is 0.053 for a split monopole geometry and 0.044 for a parabolic geometry), $\Phi_{\rm BH}$ is an absolute magnetic flux through the BH.} 
and by Tchekhovskoy et al. 2011) \footnote{
Tchekhovskoy, Narayan \& McKinney (2010) extended the magnetic power in BZ77 to high-spin BHs, (see also Tchekhovskoy, Narayan \& McKinney 2011 and Tchekhovskoy \& McKinney 2012), and obtained 
$
\dot{E}_{\rm B} = \frac{\kappa}{4\pi c} \Phi_{\rm BH}^2 \Omega_\bullet^2 f(\Omega_\bullet),
$
where $f(\Omega_\bullet) \simeq 1 + 1.38 (\Omega_\bullet r_{\rm g}/c)^2 - 9.2 (\Omega_\bullet r_{\rm g}/c)^4$ is a high-spin correction to BZ77.}. It is found that the BZ power with the formula adopted here is quite close to that given by Tchekhovkoy (2011). However, the BZ77 expression can only apply to the case with low BH spin. Similar results were also obtained by recent GRMHD numerical simulations (Nagataki 2009, 2011).

The total magnetic torque applied on the BH is (Lee et al. 2000; Li 2000; Wang et al. 2002; McKinney 2005; Lei et al. 2005; Lei \& Zhang 2011; Lei et al. 2013)
\begin{eqnarray}
T_{\rm B} = \frac{\dot{E}_{\rm B}}{\Omega_{\rm F}}=3.4 \times 10^{45} a_\bullet^2 q^{-1} m_\bullet^3 B_{\bullet,15}^2  F(a_\bullet) {\rm \ g \ cm^2 \ s^{-2}}, \nonumber \\
\end{eqnarray}
where $\Omega_{\rm F}=0.5\Omega_\bullet$ is usually taken to maximize the BZ power, and
\begin{equation}
\Omega_\bullet =\frac{a_\bullet c}{2 r_\bullet}= \frac{c^3}{G M_\bullet} \frac{a_\bullet}{2 (1+\sqrt{1-a_\bullet^2})}
\end{equation}
is the angular velocity of BH horizon.

The spin-down timescale by the BZ process can be estimated as (Lee et al. 2000; Lei et al. 2005)
\begin{equation}
t_{\rm spindown} \simeq \frac{E_{\rm rot}}{\dot{E_{\rm B}} } \simeq 2.7 \times 10^3 {\rm s} \times B_{\rm \bullet, 15}^{-2} m_\bullet^{-1} .
\label{tspindown}
\end{equation}
One can find that $t_{\rm spindown}$ is not sensitive to the initial BH spin, since both the rotational energy and spin power depend on it.

Considering a BH with an initially spin $a_{\bullet}(0)$ is slowing down by the BZ mechanism to a final spin $a_{\bullet,f}=0$. The final BH mass is then given by
\begin{equation}
M_\bullet=M_{\bullet}(0) \exp {\int_{a_{\bullet}(0)}^0 \frac{-1}{2a_\bullet-4/q} da_\bullet}.
\end{equation}
If $a_{\bullet}(0)=1$, the final BH mass will be $M_\bullet=(e^{1/4}/{\sqrt{2}})M_{\bullet}(0)=0.91 M_{\bullet}(0) $. We see that $9\%$ of the initial mass or $31\%$ of the rotational energy can be used to power GRB from the maximally rotating BH. The extracted energy is therefore less than a half of the initial rotational energy. Other energy increase the irreducible mass of the BH. For $a_{\bullet}(0)=0.5$, $M_\bullet=0.98 M_{\bullet}(0)$ or $2\%$ of the initial mass can be used to power a GRB\footnote{Atteia et al. (2017) found a maximum isotropic energy of GRBs when they studied the GRB energy distribution within redshifts $z=1-5$. Jet break measurements are needed to derive the beaming-corrected energy, which can be compared with our model predictions.}.

As the magnetic field on the BH is supported by the surrounding disk, there are some relations between $B_{\bullet}$ and $\dot{M}$. In a hyper-accreting flow in a GRB, it is possible that a magnetic flux is accumulated near the black hole horizon. Considering the balance between the magnetic pressure on the horizon and the ram pressure of the innermost part of the accretion flow (e.g. Moderski et al. 1997), one can estimate the magnetic field strength threading the BH horizon $B_{\bullet}^2 /(8\pi) = P_{\rm ram} \sim \rho c^2 \sim \dot{M} c/(4\pi r_{\bullet}^2)$, where $r_{\bullet}=(1+\sqrt{1-a_\bullet^2})r_{\rm g}$ is the radius of BH horizon. One thus has
\begin{equation}
B_{\bullet} \simeq 7.4 \times 10^{16} \dot{m}^{1/2} m_\bullet^{-1} \left(1+\sqrt{1-a_\bullet^2} \right)^{-1} \rm{G}.
\label{Bmdot}
\end{equation} 
Inserting it into Equation (\ref{eq_Lmag}), we obtain the magnetic power and torque as a function of mass accretion rate and BH spin, i.e.
\begin{eqnarray}
 \dot{E}_{\rm B} & = & 9 \times 10^{53} a_\bullet^2 \dot{m}  X(a_\bullet) \ {\rm erg \ s^{-1}} \nonumber \\
& \simeq & 1.5 \times 10^{53} a_\bullet^2 \dot{m}  \ {\rm erg \ s^{-1}},
\end{eqnarray}
\begin{eqnarray}
 T_{\rm B} & = & 1.8 \times 10^{49} a_\bullet \dot{m} m_\bullet  F(a_\bullet) \ {\rm \ g \ cm^2 \ s^{-2}} \nonumber \\
& \simeq & 1.2 \times 10^{49} a_\bullet \dot{m} m_\bullet  \ {\rm \ g \ cm^2 \ s^{-2}},
\end{eqnarray}
where $X(a_\bullet)=F(a_\bullet)/(1+\sqrt{1-a_\bullet^2} )^2$. It is found that $X(0)=1/6$, and $X(1)=\pi -2$. 

\begin{figure}[ht]
\center
\includegraphics[width=8cm,angle=0]{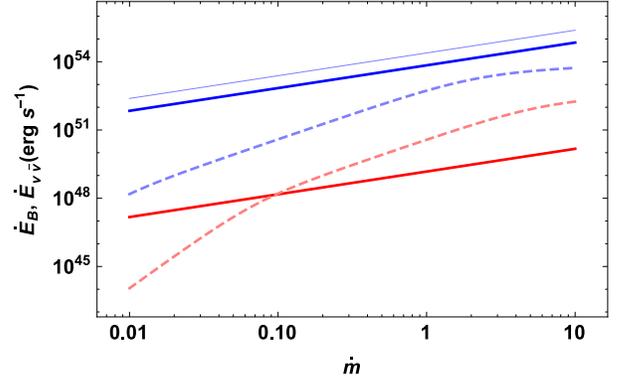}
\caption{The magnetic power $\dot{E}_{\rm B} $ as a function of accretion rate for different BH spin $a_\bullet =0.01$ (thick red solid line) and 0.99 (thick blue solid lines). The dashed lines show neutrino annihilation power $\dot{E}_{\nu\bar{\nu}}$ calculated with Equation (\ref{eq_Evv}) for $a_\bullet =0.01$ (red dashed line) and 0.99 (blue dashed line). The thin blue line is produced with the analytic expression of Tchekhovskoy, Narayan \& McKinney (2011) for $a_\bullet =0.99$, where the average magnetic flux $\langle\Phi_{\rm BH}^2/(\dot{M} r_{\rm g}^2)c\rangle^{1/2} \simeq 47$ and $\kappa=0.044$ are taken based on the numerical simulation Model A0.99f.}
\label{fig_Eb_mdot}
\end{figure}

Both neutrino annihilation and magnetic power depend on the disk mass accretion. In Figure \ref{fig_Eb_mdot}, we present the BZ power as function of accretion rate for different BH spin and compare it with the neutrino annihilation power. We find that: (1) the magnetic power is much greater than the neutrino annihilation power for a moderate to high spin BH; (2) The neutrino annihilation power dominates over the BZ power for BHs with a very small spin at high accretion rates; (3) compared with the magnetic power, $\dot{E}_{\nu \bar{\nu}}$ is much more sensitive to the mass accretion rate $\dot{m}$. Therefore, if the disk accretion rate is variable, the jet driven by the neutrino annihilation process should be highly variable. However, the MHD jet is usually subject to instabilities, such as kink instability (Wang et al. 2006) and magnetic reconnection (e.g. the Internal-Collision induced Magnetic Reconnection and Turbulence or ICMART in Zhang \& Yan 2011). These MHD processes will add complexity to the GRB lightcurves.

In the magnetic model, baryons from the disk will be suppressed by the strong magnetic field lines. For $\dot{m}>\dot{m}_{\rm ign}$, the baryon loading rate for the BZ driven jet is (Paper I),
\begin{eqnarray}
\dot{M}_{\rm j,BZ} & \simeq & 3.5 \times 10^{-7} A^{0.58} B^{-0.83} f_{\rm p,-1}^{-0.5} \theta_{\rm j,-1} \theta_{\rm B,-2}^{-1} \nonumber \\
& &  \alpha_{-1}^{0.38} \epsilon_{-1}^{0.83} \dot{m}_{-1}^{0.83} \left(\frac{m_\bullet}{3}\right)^{-0.55}  r_{z,11}^{0.5} \ M_{\sun} {\rm s}^{-1}.
\label{Eq_mdotj_BZ}
\end{eqnarray}
For $\dot{m} <\dot{m}_{\rm ign}$, the dependence of $\dot{M}_{\rm j, \nu\bar{\nu}} $ on $\dot{m}$ will be $\dot{M}_{\rm j, BZ} \propto \dot{m}^{1.9}$. In Equation (\ref{Eq_mdotj_BZ}), $f_{\rm p}$ is the fraction of protons, $r_{\rm z}$ is the distance from the BH in the jet direction, which is normalized to $10^{11}$cm. Because of the existence of a strong magnetic field, protons with an ejected direction larger than $\theta_{\rm B}$ with respect to the field lines would be blocked.

We can then define a parameter denoting the maximum available energy per baryon in the jet driven by the BZ process, 
\begin{equation}
\mu_0 \equiv \frac{\dot{E}}{\dot{M}_{\rm j,BZ} c^2} = \frac{\dot{E}_{\rm m}+\dot{E}_{\rm B}}{\dot{M}_{\rm j,BZ} c^2} = \eta (1+\sigma_0),
\label{eq:mu}
\end{equation}
where $\dot{E}_{\rm m} = \dot{E}_{\nu \bar{\nu}}+ \dot{M}_{\rm j,BZ} c^2$, and $\sigma_0 = \dot E_{\rm B}/\dot E_{\rm m}$.

The acceleration behavior of the jet is subject to uncertainties. Generally, the jet will reach a terminating Lorentz factor $\Gamma$ that satisfies
\begin{equation}
 \Gamma_{\rm min} < \Gamma < \Gamma_{\rm max},
\label{eq_gamma_bz}
\end{equation}
with the explicit value depending on the detailed dissipation process, such as kink instability (Wang et al. 2006), ICMART (Zhang \& Yan 2011) and magnetic dissipate due to the shearing interaction
between two component jets (e.g. Wang et al. 2014). In Equation (\ref{eq_gamma_bz}), $\Gamma_{\rm min}=\max(\mu_{0}^{1/3},\eta)$
($\eta=\dot{E}_{\nu \bar{\nu}} /(\dot{M}_{\rm j, BZ} c^2)$) and $\Gamma_{\rm max} =  \mu_0$, which correspond to the beginning and the end of the slow acceleration phase in a hybrid outflow, respectively (see Gao \& Zhang 2015 for a detailed discussion of the acceleration dynamics of an arbitrarily magnetized relativistic or hybrid jet).

As to the evolution of BH, we should consider both the accretion and BZ processes. The evolution equations are given by
\begin{equation}
\frac{dM_\bullet c^2}{dt} = \dot{M} c^2 E_{\rm ms} - \dot{E}_{\rm B},
\label{dMbz}
\end{equation}

\begin{equation}
\frac{dJ_\bullet}{dt} = \dot{M} L_{\rm ms} - T_{\rm B}
\label{dJbz}
\end{equation}
the evolution equation for the BH spin is then
\begin{eqnarray}
\frac{da_\bullet}{dt} = && (\dot{M} L_{\rm ms} - T_{\rm B})c/(G M_\bullet^2) - \nonumber \\
&& 2 a_\bullet (\dot{M} c^2  E_{\rm ms} - \dot{E}_{\rm B}) /(M_\bullet c^2)
\end{eqnarray}

As a BH may be spun up by accretion or spun down by the BZ mechanism, the BH spin will reach an equilibrium value when $da_\bullet/dt =0$. If the magnetic field is related to the mass accretion rate as Equation (\ref{Bmdot}), the final BH spin will be $a_\bullet^{\rm eq} \sim 0.87$.

The evolution of BH spin combining with the accretion profile will give rise to a reasonable GRB lightcurve. In addition, possible jet pression (Lei et al. 2007), episodic jet (Yuan \& Zhang 2012) and episodic accretion (by magnetic barrier, see Proga \& Zhang 2006; or by magnetically arrested disk (MAD), see Lloyd-Ronning et al. 2016) would enrich the structure of lightcurve.


\section{Prompt Emission Phase}
Now we apply the above theory to GRBs. Firstly, we study the prompt emission phase. During the this stage, the BH accretes the main part of the disk with a high accretion rate. We begin with a BH of mass $M_\bullet(0)=3 M_{\sun}$, spin $a_{\bullet}(0)$, accretion rate $\dot{M}(0)$ and with a disk of mass $M_{d}(0)$. Other parameters are taken their typical values ($r_{\rm z}=10^{11}cm, f_{\rm p}=0.1, \theta_{\rm j}=0.1, \theta_{\rm B}=0.01$).

To obtain the accretion rate profile, we adopt a simple model described in Kumar et al. (2008a, 2008b) and Metzger et al. (2008). In this model, the disk are treated as a single annulus ring with effective disk radius $r_d$, which is defined as
\begin{equation}
j(r_{\rm d}) = (G M_\bullet r_{\rm d})^{1/2} = \frac{J_{\rm d}}{M_{\rm d}}
\label{Eq:jd}
\end{equation}
where $M_{\rm d}$ and $J_{\rm d}$ are the total mass and angular momentum of the disk at time $t$. The accretion rate depends on the mass and accretion time-scale as
\begin{equation}
\dot{M} = M_{\rm d}/t_{\rm acc}
\label{Eq:dMacc}
\end{equation}
where $t_{\rm acc}=r_{\rm d}^2 / \nu \sim 2/(\alpha \Omega_K)$, and $\alpha$ is the dimensionless viscosity parameter(Shakura \& Sunyaev 1973).

The mass and angular momentum of the disk change with time as
\begin{equation}
\dot{M}_{\rm d} = - \dot{M}
\label{Eq:dMd}
\end{equation}
\begin{equation}
\dot{J}_{\rm d} = - L_{\rm ms} \dot{M}
\label{Eq:dJd}
\end{equation}

The evolutions of the BH are given by Equations (\ref{dMvv})-(\ref{dJvv}) for the neutrino annihilation model, and by Equations (\ref{dMbz})-(\ref{dJbz}) for the magnetic model. 

\begin{figure*}[ht]
\center
\includegraphics[width=7cm,angle=0]{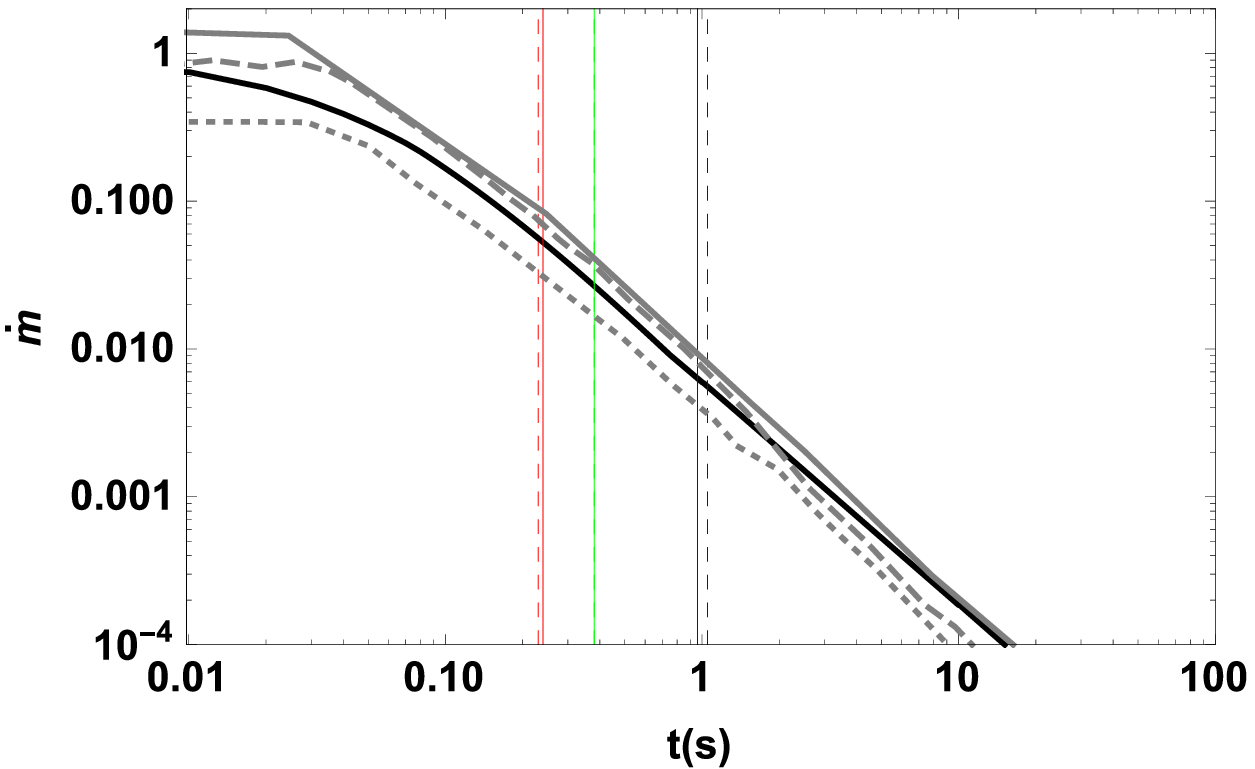}
\includegraphics[width=7cm,angle=0]{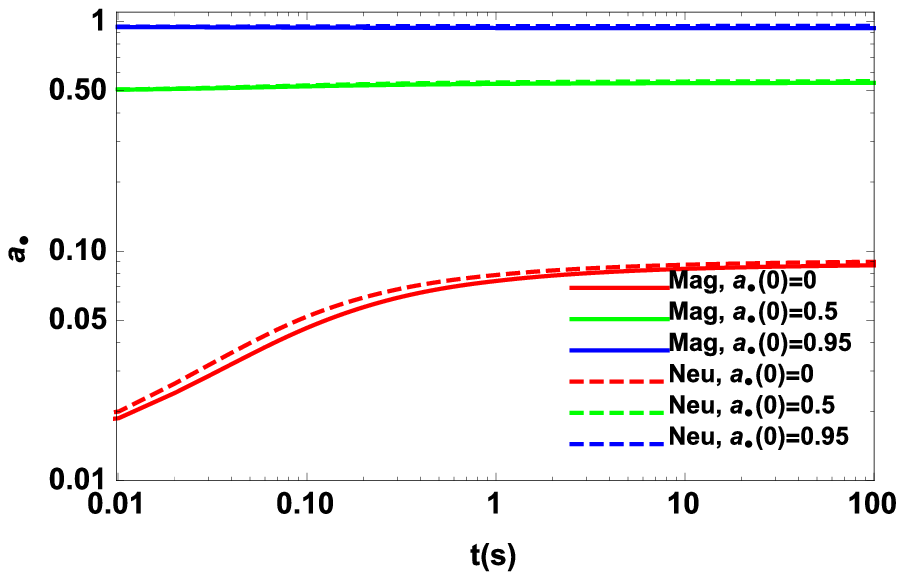}\\
\includegraphics[width=7cm,angle=0]{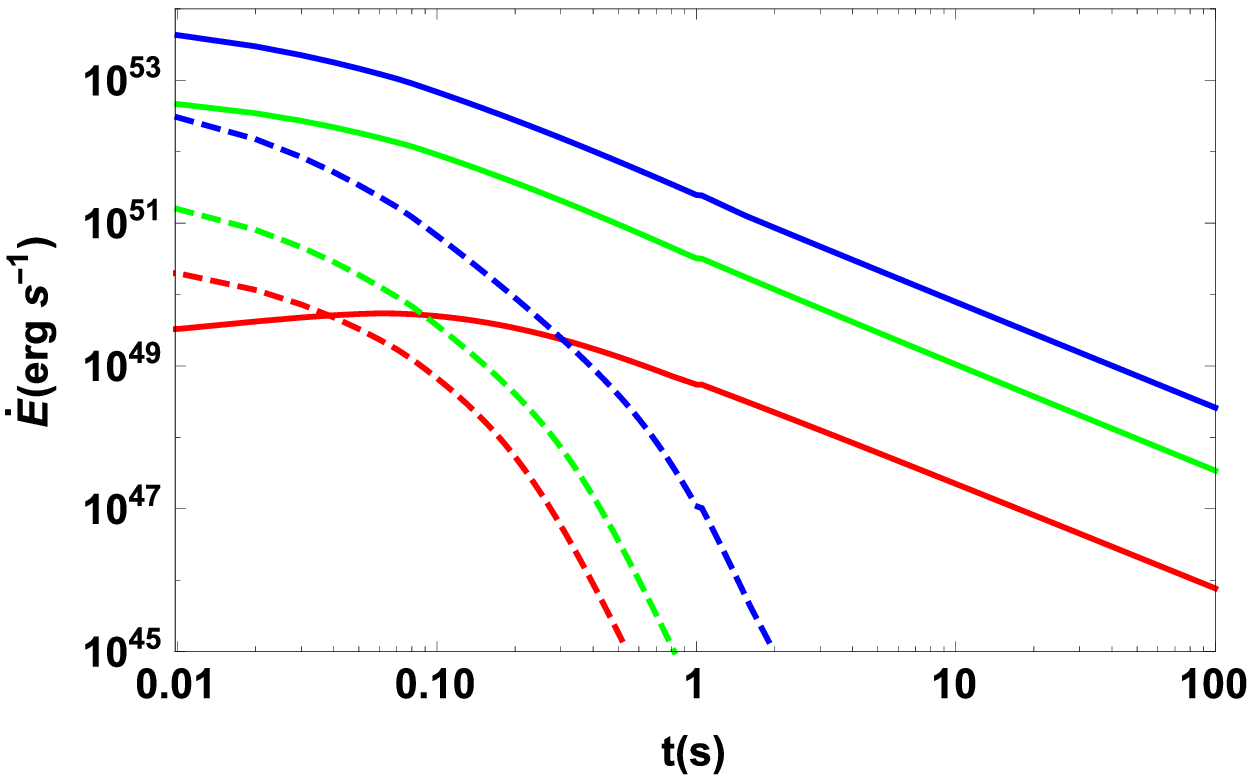} 
\includegraphics[width=7cm,angle=0]{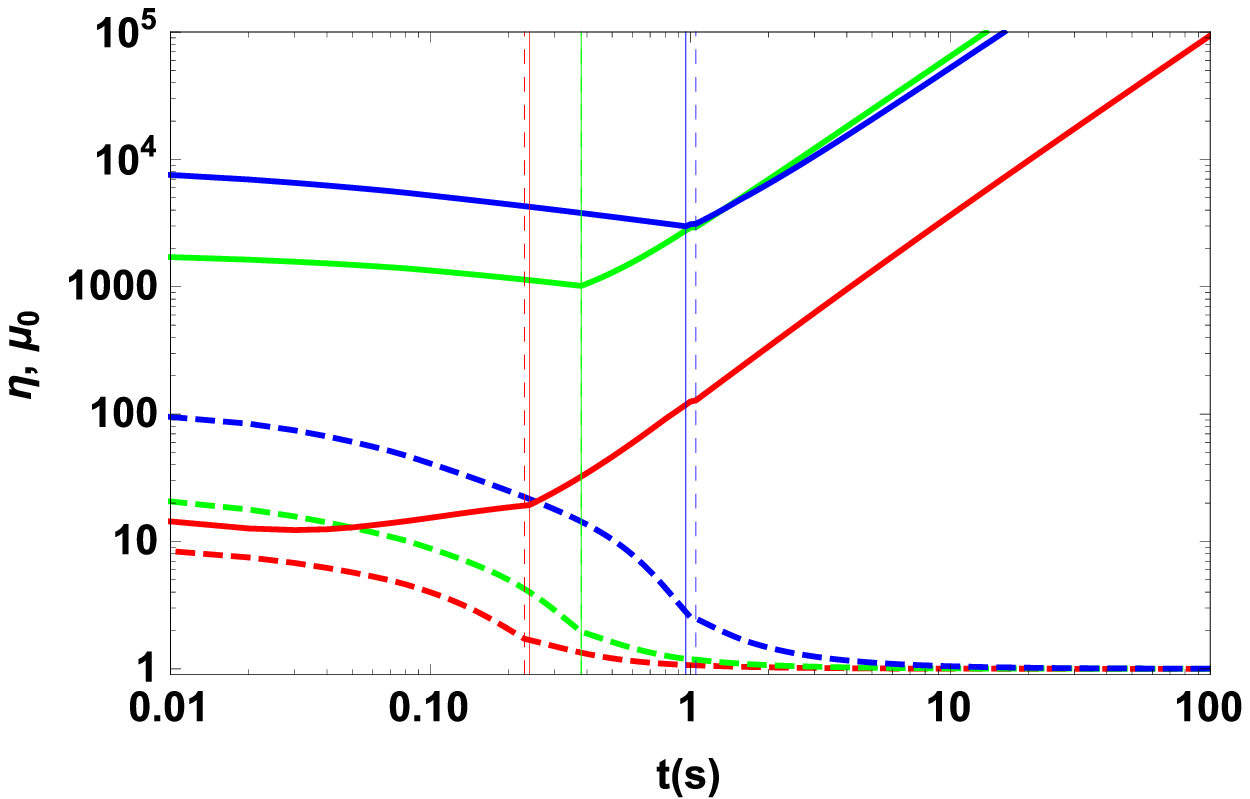} 
\caption{The time evolution of the mass accretion rate $\dot{M}$ (top left), BH spin $a_\bullet$ (top right), the jet power (lower left) and the $\eta$ ($\mu_0$) parameters (lower right). The solid lines correspond to the magnetic model and dashed lines to neutrino annihilation model. We plot three cases with different BH initial spin: $a_{\bullet,i}=0$ (red lines), 0.5 (green lines), and 0.95 (blue lines). In the left panel, the solid and dashed vertical lines mark the time when the accretion rate drops below $\dot{m}_{\rm ign}$ for neutrino and magnetic models, respectively. The igniting accretion rate $\dot{m}_{\rm ign}$ is a function of $a_\bullet$, so we have three vertical lines for each model, corresponding to different values of initial spin, i.e., $a_{\bullet}(0)=0$ (red lines), 0.5 (green lines), and 0.95 (blue lines). In the calculations, we adopt a disk mass $m_{\rm d}(0) =0.1$ and an accretion rate $\dot{m}(0) = 1$. For comparison, we also show the analytical results of fall-back rate from Rosswog (2007), which were based on numerical simulations for various NS-BH binaries with different mass ratios: $1.4:6$ (gray solid line), $1.4:14$ (gray dashed line), $1.4:16$ (gray dotted line).}
\label{fig_evl1}
\end{figure*}

\begin{figure*}[ht]
\center
\includegraphics[width=7cm,angle=0]{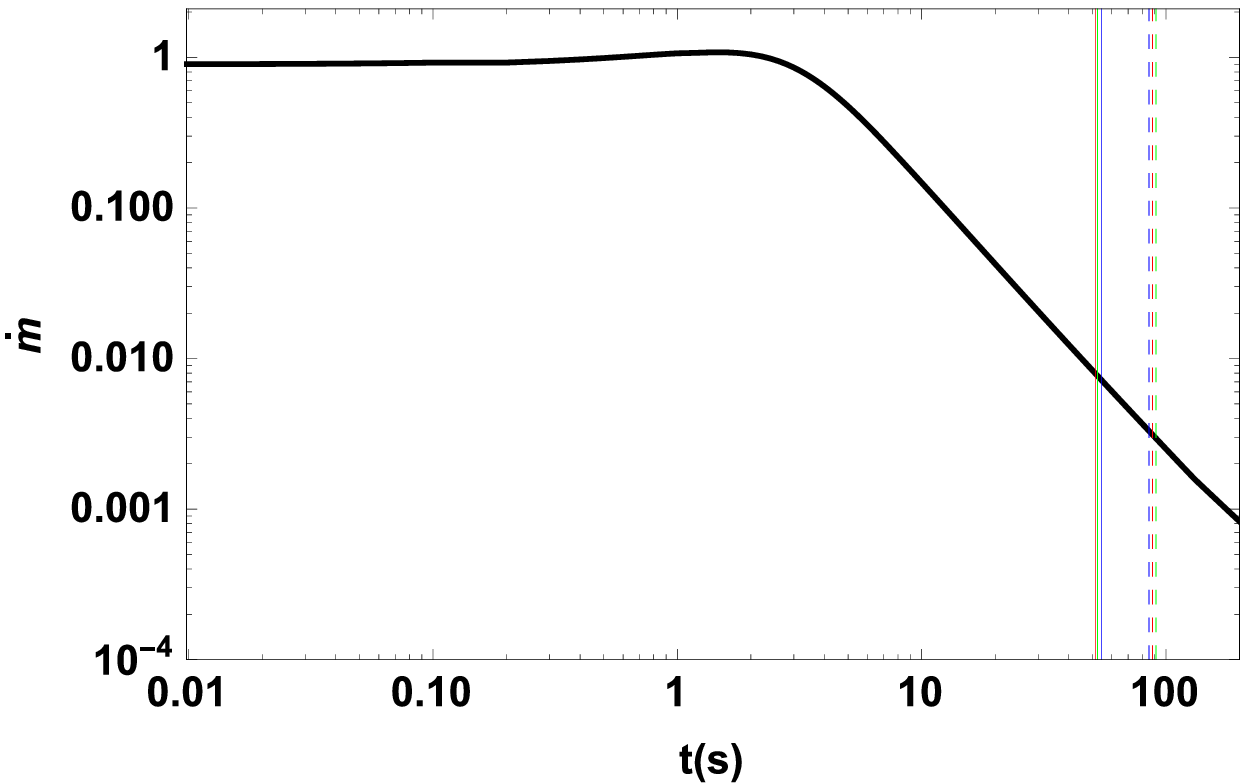}
\includegraphics[width=7cm,angle=0]{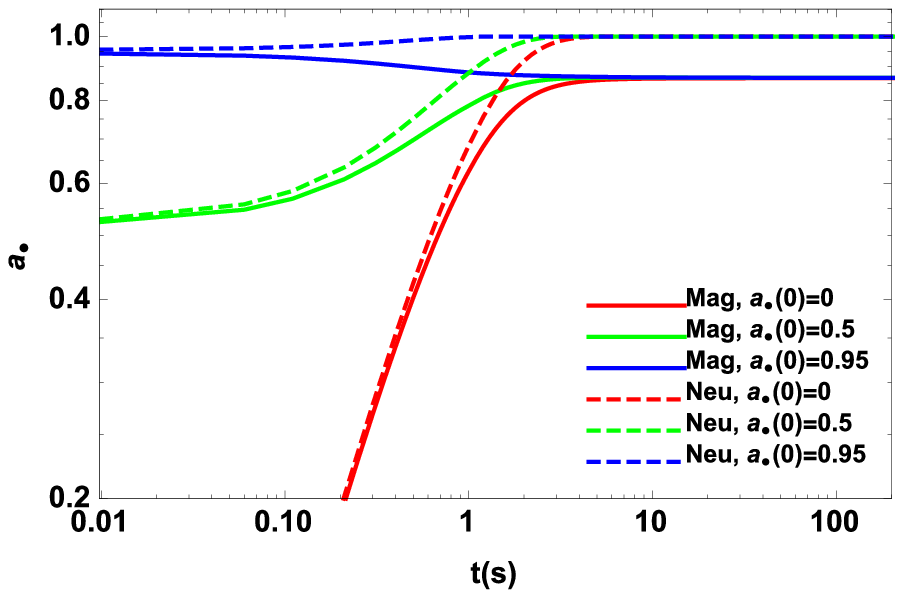}\\
\includegraphics[width=7cm,angle=0]{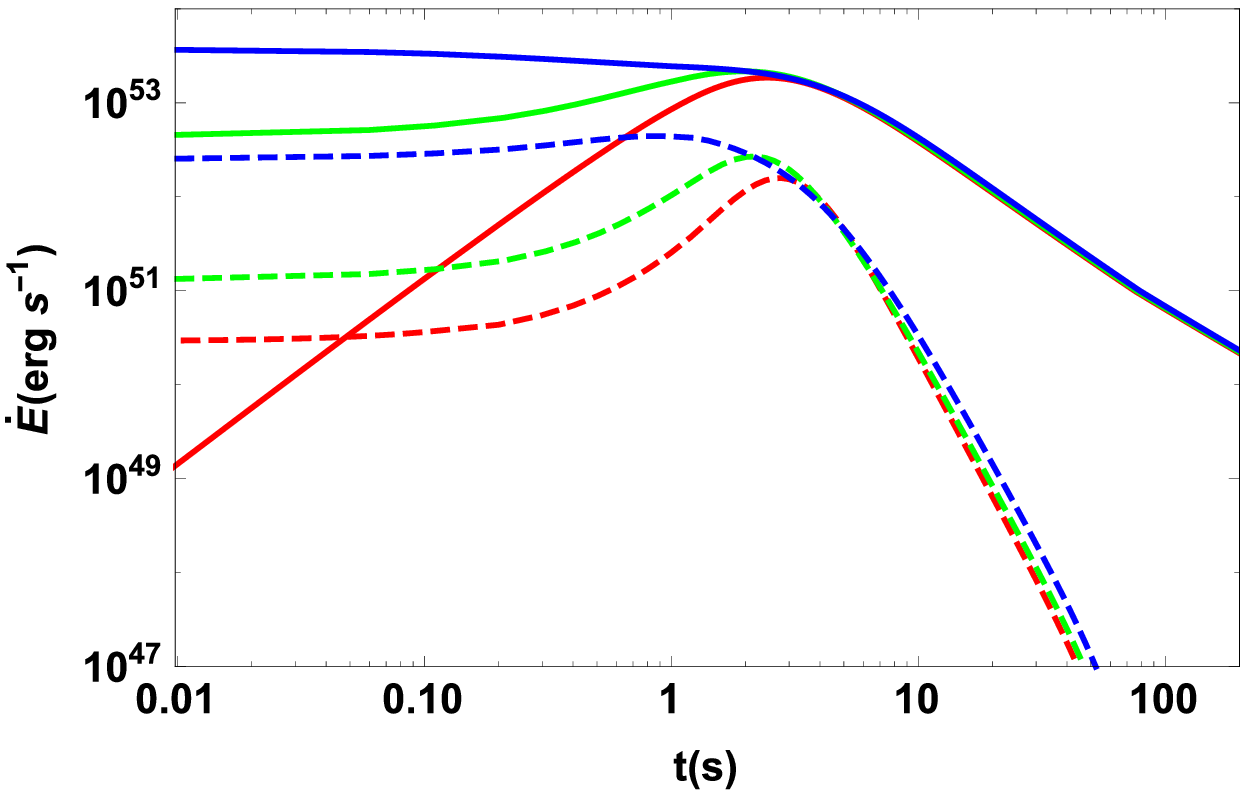} 
\includegraphics[width=7cm,angle=0]{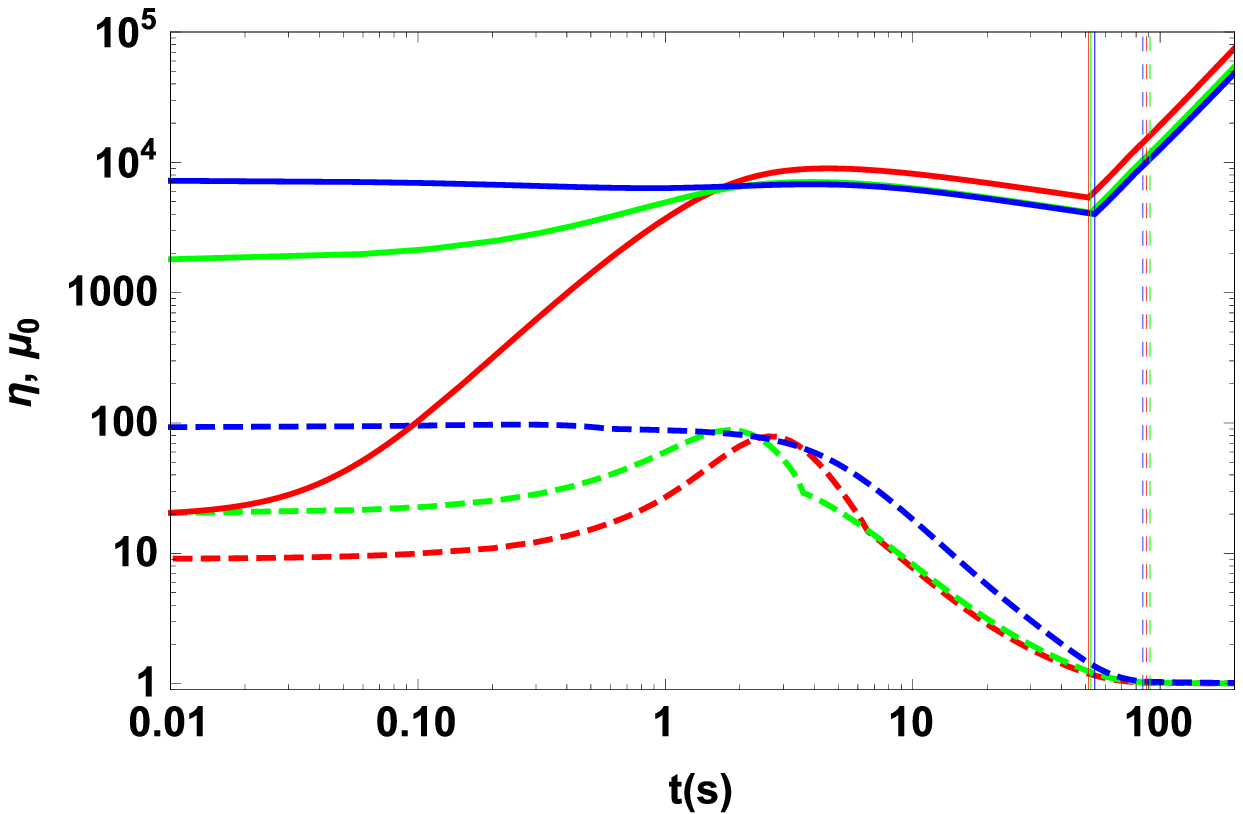} 
\caption{Same as Figure \ref{fig_evl1}, but for disk mass $m_{\rm d}(0) =10$ and accretion rate $\dot{m}(0) = 1$.}
\label{fig_evl10}
\end{figure*}

\begin{figure*}[ht]
\center
\includegraphics[width=7cm,angle=0]{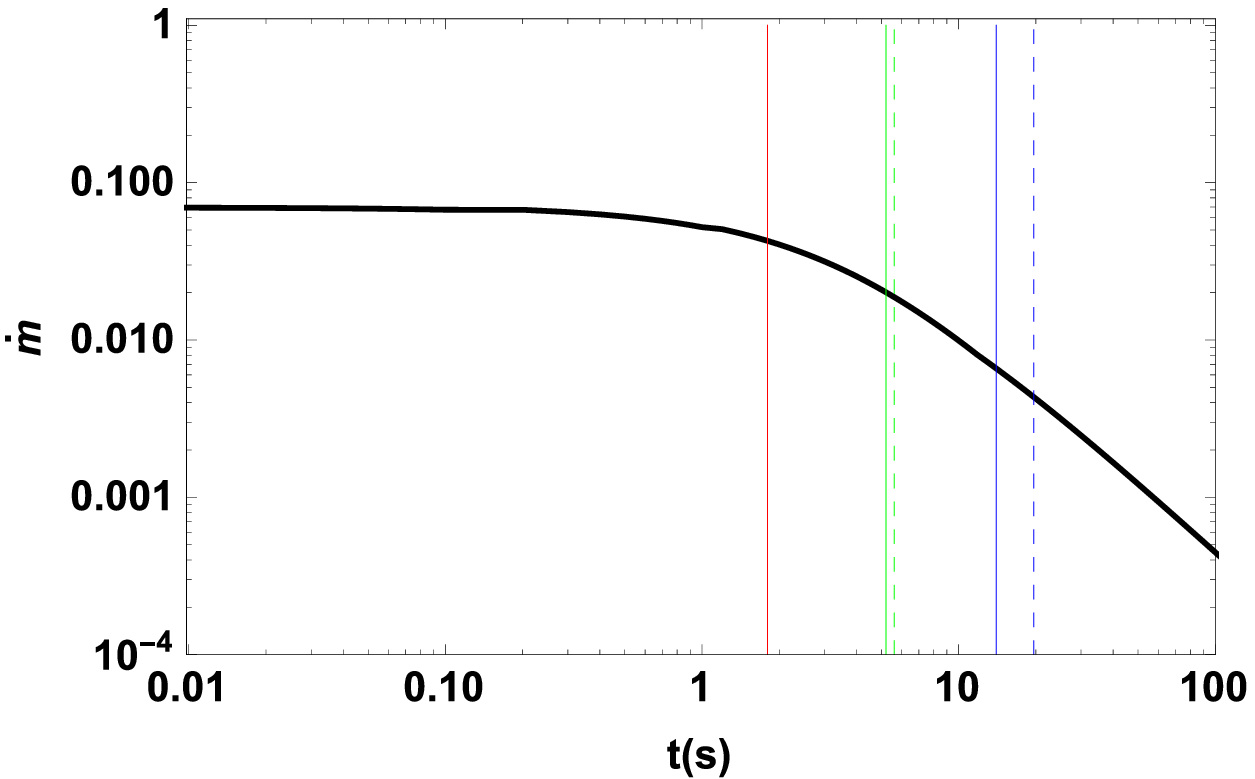}
\includegraphics[width=7cm,angle=0]{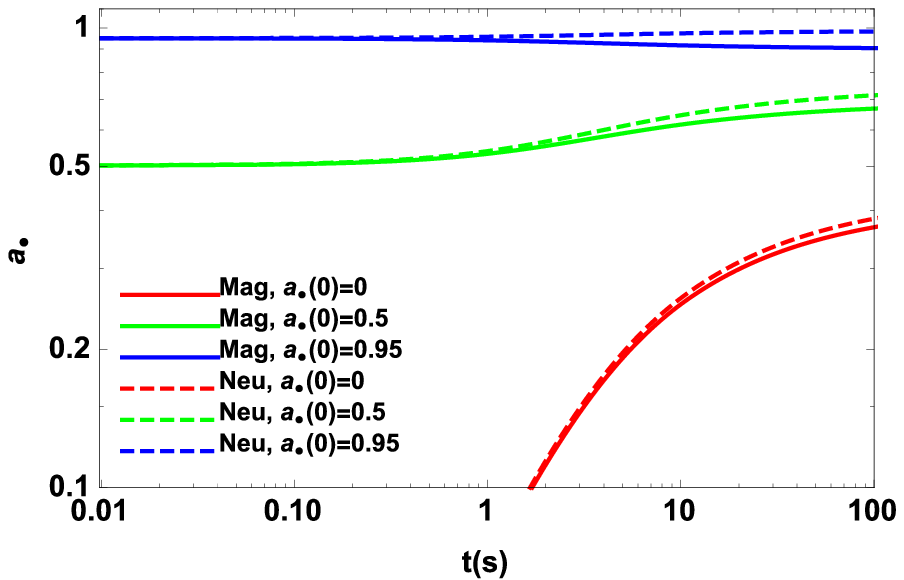}\\
\includegraphics[width=7cm,angle=0]{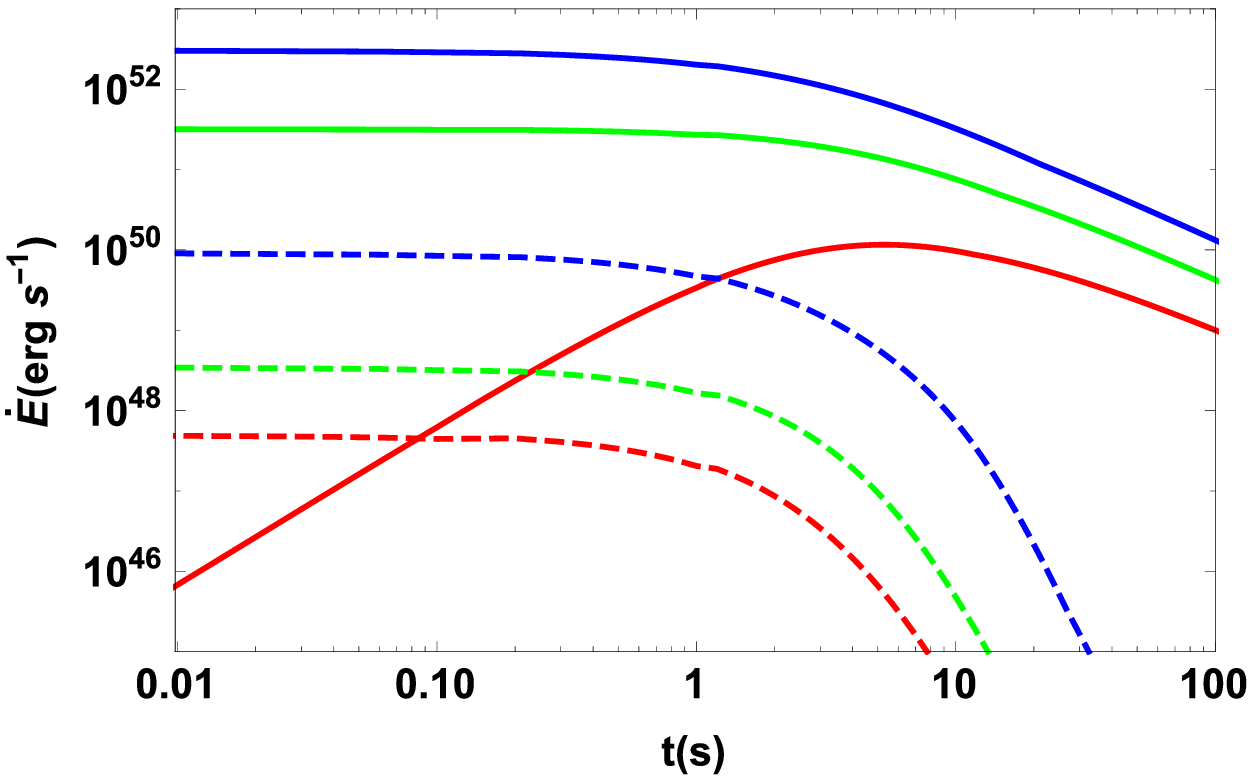} 
\includegraphics[width=7cm,angle=0]{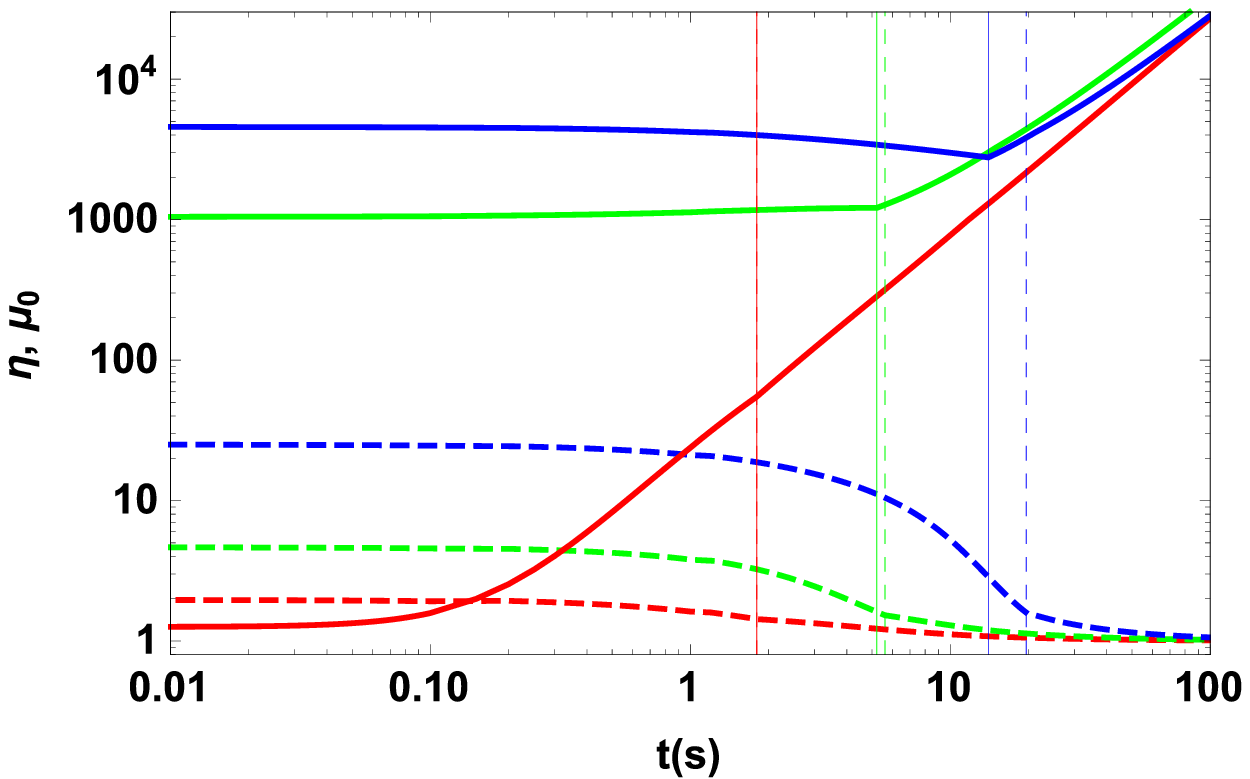} 
\caption{Same as Fig.\ref{fig_evl1}, but for disk mass $m_{\rm d}(0) =1$ and accretion rate $\dot{m}(0)= 0.1$.}
\label{fig_evl01}
\end{figure*}

Combing the evolution equations of the disk and the BH, one can get the values of $\dot{m}$, $m$ and $a_\bullet$ at each time step. With the formula obtained in Section 2, we can perform the evolution of the central engine parameters, such as $\dot{E}_{\nu \bar{\nu}}$, $\dot{E}_{\rm B}$, $\eta$ (for the neutrino model) and $\mu_0$ (for the magnetic model). The results are presented in Figures \ref{fig_evl1}-\ref{fig_evl01} for different sets of initial parameters.

Figure \ref{fig_evl1} shows the case with an initial accretion rate of $\dot{m}(0)=1$ and initial disk mass of $m_{\rm d}(0)=0.1$. The parameters of the neutrino model and the magnetic model are plotted with dashed lines and solid lines, respectively. Different colors indicate distinct initial BH spin parameters, i.e., $a_{\bullet}(0)=0$ (red lines), 0.5 (green lines), and 0.95 (blue lines). 

Top left panel exhibits the evolution of the accretion rate $\dot{m}$, which is insensitive to the BH parameters. So for the three examples exhibited in this figure, they share the same evolution curve for $\dot{m}$. The mass accretion rate decreases during the prompt phase due to angular momentum transfer. The vertical lines mark the igniting time $t_{\rm ign}$ when $\dot{m}$ becomes lower than the igniting accretion rate $\dot{m}_{\rm ign}(a)$, after which the neutrino cooling becomes unimportant. 

For the neutrino model, the BH spin is always increasing until reaching the maximum spin $\sim 0.998$ if possible (see the dashed lines in the top right panel). For the magnetic model (solid lines in the top right panel), the evolution tracks have been divided into two branches by the equilibrium spin $a_\bullet^{\rm eq}$, i.e., the increasing branch for $a_\bullet(0)<a_\bullet^{\rm eq}$ (e.g., red and blue solid lines) and the decreasing branch for $a_\bullet(0)>a_\bullet^{\rm eq}$ (e.g., the blue solid line). 

The jet power (lower left) at each time step depends on the values of accretion rate, BH spin and BH mass (the dependence on mass is weak). We find that the evolution of $\dot{E}$ generally tracts the accretion profile at late times since the evolution of the BH spin can be ignored when the majority of the disk mass is accreted. The evolution of $a_\bullet$ still has imprints on the $\dot{E}$ curve at earlier times, especially for $\dot{E}_{\rm B}$ with $a_\bullet(0)=0$ (red solid line in lower left panel). This case with lower $a_\bullet(0)=0$ is also an outlier in the three examples. Usually, we have $\dot{E}_{\rm B} > \dot{E}_{\nu \bar{\nu}}$ for all times. Only this one (the red lines) shows $\dot{E}_{\rm B} < \dot{E}_{\nu \bar{\nu}}$ at early times ($t<0.03$s). Our model, therefore, predicts that the jet composition can evolve from a thermally dominated jet to a magnetically dominated jet. Recently, the spectral study of GRB 160625B suggested a clear transition from fireball to Poynting flux dominated jet (Zhang et al. 2017), which might be an example of such a case. 

For the parameters $\eta$ and $\mu_0$ (lower right panel), the evolution path in principal follows that of the jet power $\dot{E}$ before the igniting time $t_{\rm ign}$. Actually, such tracing properties are believed to be the physics behind the empirical relation $L_{\gamma} - \Gamma_0$ (L\"u et al. 2012; Paper I; Yi et al. 2017). After $t_{\rm ign}$, the parameter $\mu_0$ begins to increase with time since the baryon loading rate drops very quickly in the BZ driven jet. For the case with $a_\bullet(0)=0.95$, we find a dip in evolution of $\mu_0$. It is worth mentioning that Gao \& Zhang (2015) found a similar feature in the temporal profile of magnetic parameter $\sigma_0$ when analysing the data of GRB 110721A.

To illustrate the effects of disk mass, we present the results of the disk with an initial accretion rate of $\dot{m}(0)=1$ but with a large initial disk mass of $m_{\rm d}(0)=10$, as shown in Figure \ref{fig_evl10}. We find that the typical duration becomes longer compared with the first example (Figure \ref{fig_evl1}) since there are more masses to be accreted by the BH. For the same reason, $t_{\rm ign}$ is also greater. The bumps in the evolution curve of $\dot{E}$ represents the competition between the effects of accretion and BH spin.

In Figure \ref{fig_evl01}, we study an example with a lower accretion rate. The duration becomes shorter because the flux is too weak to be observed at the final stage of accretion. 

The results obtained here are based on a simple analytical model. There are a number of simulations on the GRB central engine (e.g., MacFadyen \& Woosley 1999; Rosswog et al. 2003; Zhang et al. 2008; Janiuk et al. 2013; Janiuk 2017), which usually show complex behaviour of disk accretion. Direct comparisons between our results and theirs are beyond the scope of this paper. Rosswog (2007) presented an analytical model of the fall-back accretion of the bound debris based on previous 3D simulation of NS-NS and NS-BH mergers. Here, we adopt his results of the merger of NS-BH binaries with the NS mass fixed to $1.4 M_{\sun}$ and the BH mass adopted as $6 M_{\sun}$, $14 M_{\sun}$ and $16 M_{\sun}$, respectively. We estimate the fall-back accretion rate $\dot{m}_{\rm fb}$ from the fall-back accretion luminosity $L_{\rm acc}=dE_{\rm fb}/dt$ (Rosswog 2007), where $E_{\rm fb}$ denotes the difference between the potential plus kinetic energy at the start radius $r_{\rm i}$ and the potential energy at the dissipation radius $r_{\rm dis}$. Usually, the dissipation radius is taken as $r_{\rm dis}\simeq 10r_{\rm g}$ (Rosswog 2007). For comparison, we also plot the fall-back accretion rate with gray lines for the cases of different NS to BH mass ratios: $1.4:6$ (gray solid line), $1.4:14$ (gray dashed line), $1.4:16$ (gray dotted line) in Figure \ref{fig_evl1}. One can see that the evolution characteristics of the central engine presented here are generally consistent with those numerical simulation results.

\section{Late Central Engine Activities}
Many GRBs exhibit flares (Burrows et al. 2005; Chincarini et al. 2007; Falcone et al. 2007; Zhang 2007), plateaus (e.g. GRB 070110; Troja et al. 2007; Lyons et al. 2010; L\"u \& Zhang 2014; L\"u et al. 2015; Gao et al. 2016a; Li et al. 2016; Chen et al. 2017), or giant bumps (e.g. GRB 121027A and GRB 111209A; Wu et al. 2013; Gao et al. 2016b) in X-ray lightcurves. These observations suggest that the GRB central engine is long-lived. Various models are invoked to interpret these activities, such as continuous energy injection from the spindown power of a magnetar, and the re-start of accretion onto a BH. 

Here, we consider the BH central engine with fall-back accretion. The evolution of the fall-back accretion rate are described with a broken-power-law function of time as (Chevalier 1989; MacFadyen et al. 2001; Zhang et al. 2008; Dai \& Liu 2012)

\begin{eqnarray}
\dot{M}_{\rm fb} = \dot{M}_{\rm p} \left[ \frac{1}{2}\left(\frac{t-t_0}{t_{\rm p}-t_0} \right)^{-1/2} +  \frac{1}{2}\left(\frac{t-t_0}{t_{\rm p}-t_0} \right)^{5/3} \right]^{-1},
\label{dotm}
\end{eqnarray}
where $t_0$ is the beginning time of the fall-back accretion in the local frame. 

As an example, we assume a fall-back accretion starting at $t_0=1000$s, reaching the peak at $t_{\rm p}=1500$s. The peak accretion is adopted as $\dot{M}_{\rm p}=10^{-4} M_\sun s^{-1}$. Since $\dot{M}_{\rm p}$ is far below the igniting accretion rate, the neutrino annihilation power cannot explain the late time X-ray activities observed in both short and long GRBs (Fan et al. 2005). We ignore the contribution from neutrino annihilations, and assume that the jet is powered by the BZ process in the calculations. The baryon loading in this stage is quite uncertain since neutrino cooling is shut off and a strong wind kicks in, one cannot make robust predictions. In this paper, we do not calculate baryon loading and the parameter $\mu_0$ during the late BH central engine activity phase, although the jet is expected dirtier due to the strong disk wind expected in an advection dominated accretion flow (ADAF).

\begin{figure*}[ht]
\center
\includegraphics[width=7cm,angle=0]{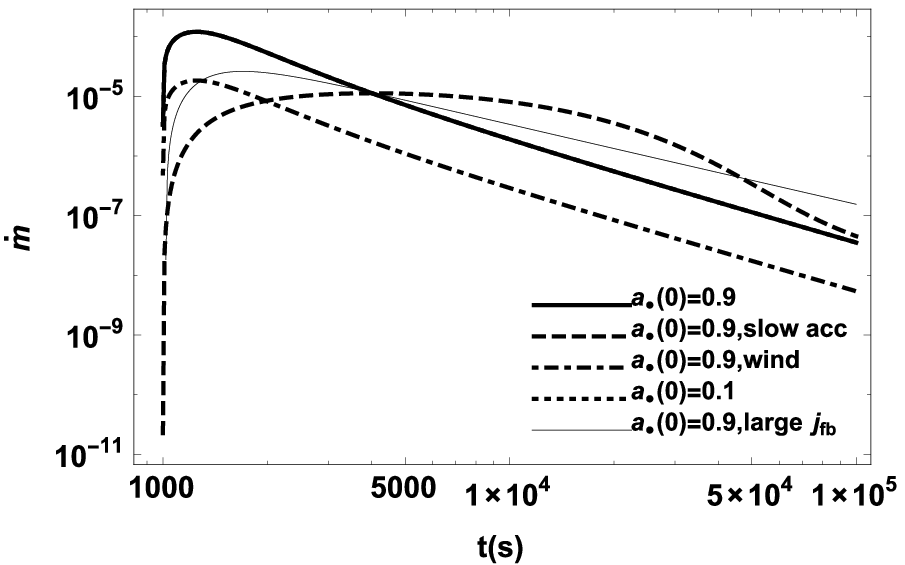}
\includegraphics[width=7cm,angle=0]{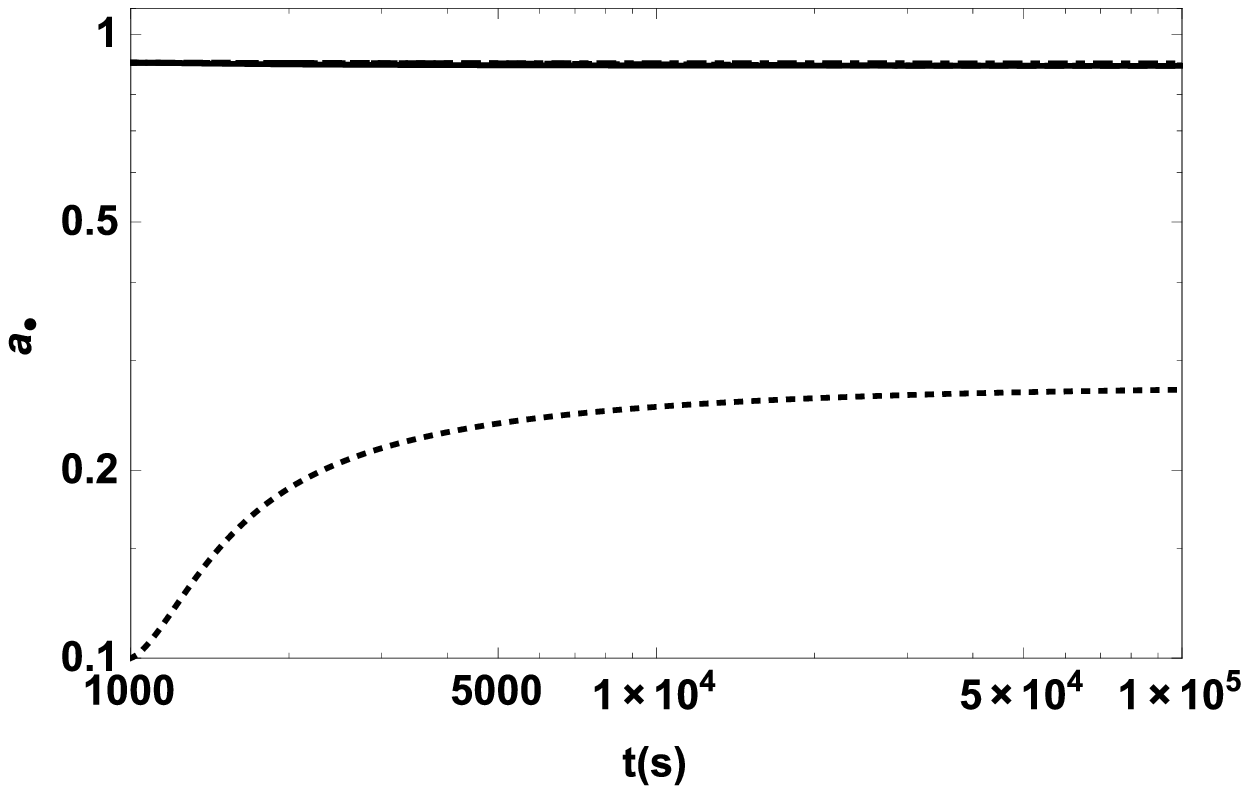}\\
\includegraphics[width=7cm,angle=0]{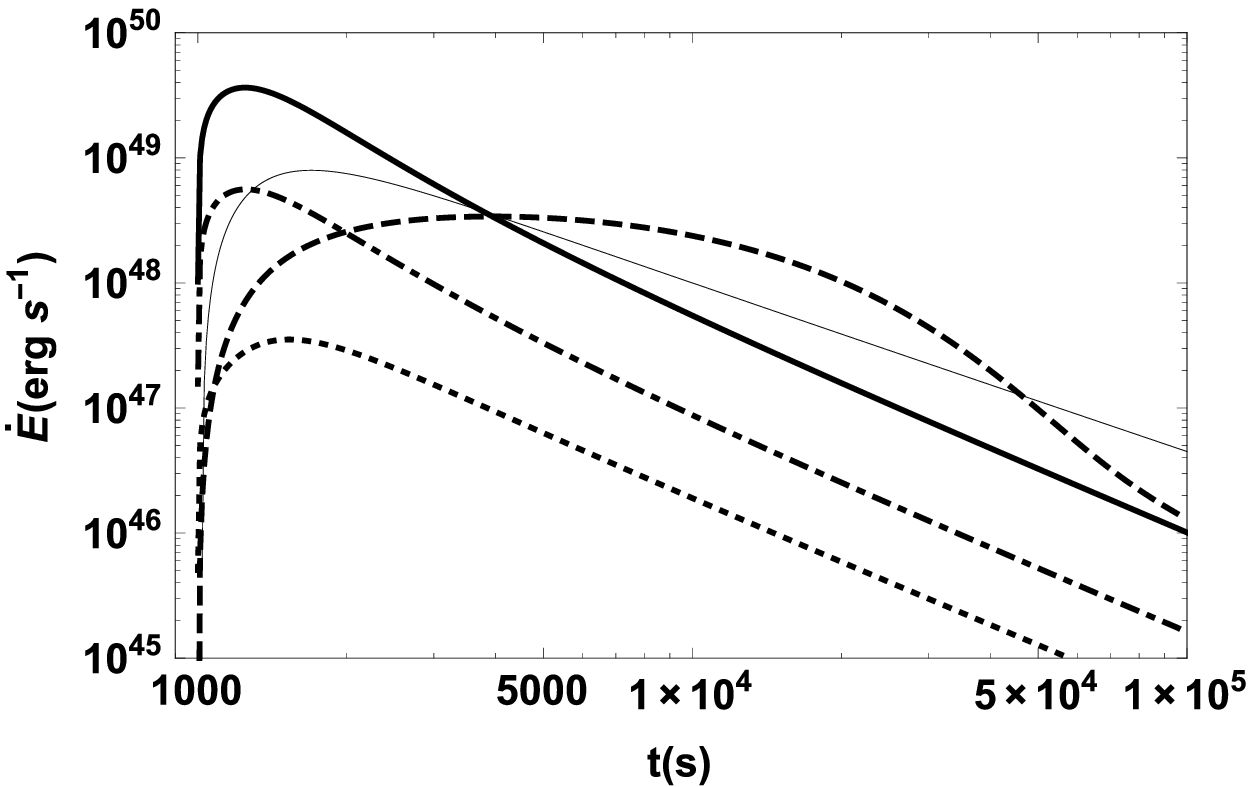} 
\caption{The evolutions of the accretion rate (top left), BH spin (top right) and BZ jet power (bottom) as a function of time during late central engine activities. We study four different models, model I (thick solid lines): $a_\bullet(0)=0.9$, rapid accretion $\dot{M}=\dot{M}_{\rm fb}$; model II (dashed lines): $a_\bullet(0)=0.9$, slow accretion with viscosity timescale of $\tau_{\rm vis}=10^4 s$; model III (dot-dashed lines): $a_\bullet(0)=0.9$, rapid accretion, but with disk outflow, $s=0.5$; model IV (dotted lines): same as model I but with small BH spin $a_\bullet(0)=0.1$; model V (thin solid lines): same as model I but assuming that the specific angular momentum $j_{\rm fb}$ of the fall-back gas is large ($f_\Omega =0.4$).}
\label{fig_evl_late}
\end{figure*}

First, we present the results of a fall-back accretion disk with rapid accretion surrounding a fast spinning BH $a_\bullet(0)=0.9$ (model I, thick solid lines in Figure \ref{fig_evl_late}). The BH accretion just follows the fall-back rate, i.e., $\dot{M}=\dot{M}_{\rm fb}$. As shown in  Figure \ref{fig_evl_late}, there is a weak evolution in the BH spin for this case. We find that the evolution of jet power just tracks that of the fall-back accretion rate.

If the viscosity parameter $\alpha$ is too small, the disk will undergo very slow accretion. We introduce a large viscosity timescale $\tau_{\rm vis}$ to model the slow accretion. The accretion rate onto the BH can be estimated as
\begin{equation}
\dot{M} = \frac{1}{\tau_{\rm vis}} e^{-t/\tau_{\rm vis}} \int_{t_0}^t e^{t^\prime /\tau_{\rm vis}} \dot{M}_{\rm fb} dt^\prime.
\end{equation} 
We, therefore, in the second case (model II) take $\tau_{\rm vis}=10000$s. The results are presented with dashed lines in Figure \ref{fig_evl_late}. The accretion rate becomes flat until $\tau_{\rm vis}$ and then begins to decay. Interestingly, we find a plateau in the jet power evolution.

Since the main part of the disk is already accreted, the mass accretion rate in this afterglow stage is very small. The disk will be dominated by advection. The feature of an adevction dominated disk is that it has a strong wind which is driven by a positive Bernouilli constant (Narayan and Yi 1994). Recently, Mu et al. (2016) took into account the effects of outflow in the accretion disk when interpreting X-ray flares. 
Due to the existence of mass loss into the wind, the accretion rate is expected to decrease inward in a scaling form 
\begin{equation}
\dot{M} \simeq \dot{M}_{\rm fb,r_d} \left(\frac{r_{\rm ms}}{r_{\rm d}}\right)^s, 
\label{eqMdotaccr}
\end{equation}
where $0 \leq s \leq 1$, $r_{\rm d}$ is the outer edger of the disk. We therefore consider a disk with $r_d=100 r_g$, $a_\bullet(0)=0.9$, and $s=0.5$, as shown with dot-dashed lines (model III) in Figure \ref{fig_evl_late}. The accretion rate and jet power have been significantly suppressed by the outflow. 

To check the effects of BH spin, we change the parameter $a_\bullet(0)=0.1$, as shown with dotted lines (model IV) in Figure \ref{fig_evl_late}. Other parameters are the same as the first case. We find a strong evolution in BH spin, and the peak of the jet power tends to be flatter than the case with a high BH spin.

Finally, the fall-back material may contain large specific angular momentum, which in turn will shape the profile of accretion rate. Supposing that the gas at fall-back radius $r$ has an angular velocity $\Omega$ equal to a fraction $f_\Omega$ of the local Keplerian angular velocity $\Omega_{\rm k} (r)$, the specific angular momentum of this gas can then be written as (Kumar et al. 2008b)
\begin{equation}
j_{\rm fb} \simeq 3.8 \times 10^{18} m_{\bullet,1}^{1/2} r_{10}^{1/2} f_\Omega(r) {\rm cm^2 s^{-1} }. 
\end{equation}
where $r_{10}=r/10^{10}\rm cm$ and $m_{\bullet,1}=m_\bullet/10$. The gas at $r$ will fall to the disk at a time around the fall-back time $t\sim t_{\rm fb} \simeq 2 (r^3/G M_\bullet)^{1/2}$. One  therefore finds that the specific angular momentum increases with time as $j_{\rm fb} \simeq 4.7 \times 10^{18}  t_2^{1/3} m_{\bullet, 1}^{2/3} f_\Omega(r) \ {\rm cm^2 s^{-1} }$. The evolution of the disk can be described with a model adopted in Kumar et al. (2008a, 2008b),
\begin{eqnarray}
\dot{M}_{\rm d} = \dot{M}_{\rm fb} - \dot{M}, \nonumber \\
\dot{J}_{\rm d} = j_{\rm fb} \dot{M}_{\rm fb} - L_{\rm ms} \dot{M},
\end{eqnarray}  
where the accretion rate $\dot{M}$ is estimated with Equation (\ref{Eq:dMacc}). In Figure \ref{fig_evl_late}, we present the results of a BH-fall-back disk system with $a_\bullet(0)=0.9$ and $f_\Omega=0.4$ (model V, thin solid lines). Since the angular momentum determines the fall-back radius (see Equation (\ref{Eq:jd})) and $t_{\rm acc}\sim 2/(\alpha \Omega_{\rm K}(r_{\rm d}))$, the large angular momentum of the fall-back material leads to a longer accretion time $t_{\rm acc}$ and thereby a shallower lightcurve.

\section{Discussions}
The central engine of GRBs is likely a hyperaccreting BH. The neutrino annihilation and BZ processes are two candidate mechanisms for powering GRB jets. In this paper, we obtained the analytical solutions to the neutrino and magnetic models, and studied the time evolution of the central engine parameters for these two models. 

The evolution of accretion rate and BH spin have strong effects on the evolution of central engine parameters such as, $\dot{E}$, $\eta$ and $\mu_0$. The neutrino annihilation power is generally weaker that the BZ power. It fails to produce the long term X-ray activities observed in many GRBs. The magnetic model remains the leading candidate mechanism to interpret the X-ray flares, giant bumps and plateaus. For a BH central engine with small initial spin $a_\bullet(0)$, the jet may be first dominated by the neutrino annihilation power and then by the BZ power, leading to a transition from a thermally-dominated fireball to a Poynting flux dominated flow as is observed in some GRBs, e.g. GRB 160625B. 

There are several predictions in our model, such as the transition from a thermal to a magnetic dominated jet, the evolution of $\mu_0$, and the late time plateaus. Systematic comparisons of these predictions against a large GRB sample are needed to test the BH central engine models. Some examples (e.g., GRB 160625B and GRB 110721A) have been observed that are consistent with our model predictions.  

This work focuses on the BH-accretion central engine models. Metzger et al. (2011) and Beniamini et al. (2017) performed detailed investigations on the magnetar central engine model for GRBs. The comparison between these two models is desirable. In principle, the BH central engine is more complex, which contains two energy mechanisms (the neutrino annihilation and BZ processes) and two systems (the BH and disk). To predict a lightcurve, one needs to consider the evolution of both the central BH and the surrounding disk. Due to these intrinsic differences, our results show unique predictions on temporal evolutions of $\dot{E}$ and $\mu_0$, especially for the case with a small $a_\bullet(0)$. we hope our results can be used to distinguish the BH model from the magnetar model with observational data. 

In this paper, we ignore the baryon loading during the late time central engine activities, since there is no good knowledge on the thermally driven wind at low accretion rates when neutrino cooling totally shuts off. Our analytical solutions are based on the numerical results of a standard NDAF model. We did not include the effects, such as magnetic coupling (Lei et al. 2009), inner boundary torque (Xie et al. 2016) and vertical structure (Liu et al. 2014). These effects may be important, but usually depend on some uncertain parameters. GRMHD simulations will help to give a better understanding of these issues.

\acknowledgments
We thank H. Gao and Q. Yuan for helpful discussions. The Numerical calculations were performed by using a high performance computing cluster (Hyperion) of HUST. This work is supported by the National Basic Research Program ('973' Program) of China (grants 2014CB845800), the National Natural Science Foundation of China (U1431124, 11773010, 11673068, 11433009, 11603006, 11533003 and U1731239). X.F. W also acknowledges the support by the Key Research Program of Frontier Sciences (QYZDB-SSW-SYS005), the Strategic Priority Research Program “Multi-waveband Gravitational Wave Universe” (Grant No. XDB23000000) of the Chinese Academy of Sciences. E.W. L also acknowledges support by the Guangxi Science Foundation (2016GXNSFCB380005) and special funding for Guangxi distinguished professors (Bagui Yingcai \& Bagui Xuezhe).



\begin{thebibliography}{99}
\bibitem[{Abdo et al.}(2009)]{Abdo09} Abdo, A. A., Ackermann, M., Arimoto, M., et al. 2009, Sci, 323, 1688
\bibitem[{Abramowicz et al.}(1988)]{Abramowicz88} Abramowicz, M., Czerny, B., Lasota, J. P., \& Szuszkiewicz, E. 1988, ApJ,332, 646
\bibitem[{Atteia et al.}(2107)]{Atteia17} Atteia, J.-L., Heussaff, V., Dezalay, J.-P., et al. 2017, ApJ, 837, 119
\bibitem[{Bardeen et al.}(1972)]{Bardeen72} Bardeen, J. M., Press, W. H., \& Teukolsky, S. A. 1972, ApJ, 178, 347
\bibitem[{Begelman}(1978)]{Begelman78} Begelman, M. C. 1978, MNRAS, 184, 53
\bibitem[{Beniamini et al.}(2017)]{Beni17} Beniamini, P., Giannios, D., Metzger, B. D., arXiv: 1706.05014
\bibitem[{Blandford \& Znajek}(1977)]{BZ77} Blandford, R. D., \& Znajek, R. L. 1977, MNRAS, 179, 433 (BZ77)
\bibitem[Burrows et al. (2005)]{B05} Burrows D.~N., Romano P., Falcone A., et al. 2005, Science, 309, 1833\bibitem[{Chen et al.}(2017)]{Chen17} Chen, W., Xie, W., Lei, W. H., Zou, Y. C., L\"{u}, H. J., Liang, E. W., Gao, He., \& Wang, D. X. 2017,  accepted for publication in ApJ, arXiv:1709.08285
\bibitem[{Chen \& Beloborodov}(2007)]{CB07} Chen, W. X., \& Beloborodov, A. M. 2007, ApJ, 657, 383
\bibitem[{Chevalier}(1989)]{Chevalier89} Chevalier, R.~A.\ 1989, \apj, 346, 847
\bibitem[{Chincarini et al.}(2007)]{C07} Chincarini, G., Morettti, A., Romano, P., et al. 2007, ApJ, 671, 1903
\bibitem[{Dai \& Liu}(2012)]{Dai12} Dai, Z.~G., \& Liu R.-Y.\ 2012, \apj, 759, 58
\bibitem[{Di Matteo et al.}(2002)]{DPN02} Di Matteo, T., Perna, R., {\&} Narayan, R. 2002, ApJ, 579, 706 (DPN02)
\bibitem[{Eichler et al.}(1989)]{Eichler89} Eichler, D., Livio, M., Piran, T., \& Schramm, D. N. 1989, Nature, 340, 126
\bibitem[{Falcone et al.}(2007)]{F06} Falcone, A. D., et al. 2007, ApJ, 671, 1921
\bibitem[{Fan et al.}(2005)]{Fan05} Fan, Y. Z., Zhang, B., \& Proga, D. 2005, ApJ, 635, L129
\bibitem[{Fryer et al.}(1999)]{Fryer99} Fryer, C. L., Woosley, S. E., Herant, M., \& Davies, M. B. 1999, ApJ, 520,650
\bibitem[{Gao \& Zhang}(2015)]{GZ15}Gao, H., Zhang, B. 2015, ApJ, 801, 103
\bibitem[Gao et al.(2016a)]{Gao16a} Gao, H., Lei, W.~H., You, Z.~Q., \& Xie, W. 2016a, ApJ, 826, 141
\bibitem[Gao et al.(2016b)]{Gao16b} Gao, H., Zhang, B., \& L\"{u}, H.~J. 2016b, Phys. Rev. D 93, 044065
\bibitem[{Ghirlanda et al.}(2012)]{Gld12} Ghirlanda, G., Nava, L., Ghisellini, G., et al. 2012, MNRAS, 420, 483
\bibitem[{Gu et al.}(2006)]{Gu06} Gu, W. M., Liu, T., {\&} Lu, J. F., 2006, ApJ, 643, L87
\bibitem[{Janiuk et al.}(2004)]{Janiuk04} Janiuk, A., Perna, R., Di Matteo, T., \& Czerny, B. 2004, MNRAS, 355, 950
\bibitem[{Januik et al.}(2007)]{Januik07} Januik, A., Yuan, Y., Perna, R., {\&} Di Matteo, T. 2007, ApJ, 664, 1011
\bibitem[{Januik \& Yuan }(2010)]{Januik10} Janiuk, A., \& Yuan, Y. 2010, A\&A, 509, 55
\bibitem[{Januik et al.}(2013)]{Januik13} Januik, A., Mioduszewski, P., \& Moscibrodzka, M. 2013, ApJ, 776, 105
\bibitem[{Januik}(2017)]{Januik17} Januik, A. 2017, ApJ, 837, 39
\bibitem[{Katz}(1977)]{Katz97} Katz, J. 1977, ApJ, 215, 265
\bibitem[{Kawanaka et al.}(2013)]{K13} Kawanaka, N., Piran, T., \&  Krolik, J. H. 2013, ApJ, 766, 31
\bibitem[{Kohri \& Mineshige}(2002)]{Kohri02} Kohri, K., \& Mineshige, S. 2002, ApJ, 577, 311
\bibitem[{Kohri et al.}(2005)]{Kohri05} Kohri, K., Narayan, R., \& Piran, T. 2005, ApJ, 629, 341
\bibitem[{Kumar et al.}(2008a)]{Kumar08a} Kumar, P., Narayan, R., \& Johnson,J. L. 2008a, MNRAS, 388, 1729
\bibitem[{Kumar et al.}(2008b)]{Kumar08b} Kumar, P., Narayan, R., \& Johnson,J. L. 2008b, Science, 321, 376
\bibitem[{Lei et al.}(2005)]{Lei05} Lei, W. H., Wang, D. X., \& Ma, R. Y. 2005, ApJ, 619, 420
\bibitem[{Lei et al.}(2007)]{Lei07} Lei, W. H., Wang, D. X., Gong, B. P., \& Huang, C. Y. 2007, A\&A, 468, 563 
\bibitem[{Lei et al.}(2009)]{Lei09} Lei,W. H., Wang,D. X. , Zhang, L., Gan,Z. M., Zou, Y. C. \& Xie, Y. 2009, ApJ, 700, 1970
\bibitem[{Lei et al.}(2010)]{Lei10} Lei,W. H., Wang,D. X. , Zhang, L., Gan,Z. M. \&  Zou, Y. C. 2010, Sciences in China (G), 2010, 53(s1), 98
\bibitem[Lei \& Zhang(2011)]{Lei11} Lei, W.~H., \& Zhang, B.\ 2011, \apjl, 740, L27
\bibitem[{Lei et al.}(2013)]{Lei13} Lei, W. H., Zhang, B. \& Liang, E. W. 2013, ApJ, 756, 125 (Paper I)
\bibitem[{Lee et al.}(2000)]{Lee00} Lee, H. K., Wijers, R. A. M. J., \& Brown, G.E. 2000, Physics Reports, 325, 83
\bibitem[Li(2000)]{Li00} Li, L.~X.\ 2000, \prd, 61, 084016
\bibitem[Li et al.(2016)]{Li16} Li, A., Zhang, B., Zhang, N.~B., Gao, H., Qi, B., Liu, T. 2016, Phys. Rev. D, 94, 083010
\bibitem[{Liang et al.}(2010)]{Liang10} Liang, E.-W., Yi, S.-X., Zhang, J., et al. 2010, ApJ, 725, 2209
\bibitem[{Liang et al.}(2015)]{Liang15} Liang, E. W., Lin, T. T., L\"{u}, J., et al. 2015, ApJ, 813, 116
\bibitem[{Liu et al.}(2007)]{Liu07} Liu, T., Gu, W. M., Xue, L., {\&} Lu, J. F. 2007, ApJ, 661, 1025
\bibitem[{Liu et al.}(2014)]{Liu14} Liu, T., Yu, X. F., Gu, W. M., \& Lu, J. F. 2014, ApJL, 791, 69
\bibitem[{Liu et al.}(2015)]{Liu15} Liu, T., Hou, S. J., Xue, L., \& Gu, W. M. 2015, ApJS, 218, 12
\bibitem[{Liu et al.}(2017)]{Liu17} Liu, T., Gu, W. M., {\&} Zhang, B. 2017, New Astronomy Review, in press (arXiv:1705.05516)
\bibitem[{Lloyd-Ronning et al.}(2016)]{L16} Lloyd-Ronning, N. M., Dolence, J. C., \& Fryer, C. L. 2016, MNRAS, 461, 1045
\bibitem[{L\"{u} \& Zhang}(2014)]{LZ14} L\"{u}, H.~J., \& Zhang, B. 2014, ApJ, 785, 74
\bibitem[{L\"{u} et al.}(2015)]{Lv15} L\"{u}, H.~J., Zhang, B., Lei, W.~H., Li, Y., Lasky, P.~D., 2015, ApJ, 805, 89
\bibitem[{L\"{u} et al}(2012)]{LZ12} L\"{u}, J., Zou, Y. C., Lei, W. H., et al. 2012, ApJ, 751, 49
\bibitem[Lyons et al. (2010)]{L10} Lyons N., O'Brien P. T., Zhang B., et al. 2010, MNRAS, 402,705
\bibitem[{MacFadyen \& Woosley}(1999)]{MacFadyen99} MacFadyen, A. I., \& Woosley, S. E. 1999, ApJ, 524, 262
\bibitem[MacFadyen et al.(2001)]{M01} MacFadyen, A.~I., Woosley, S.~E., \& Heger, A.\ 2001, \apj, 550, 410
\bibitem[McKinney(2005)]{Mckinney05} McKinney, J.~C.\ 2005, \apjl, 630, L5
\bibitem[{Meszaros \& Rees}(1997)]{MR97} M\'{e}sz\'{a}ros, P., \& Rees, M. J. 1997, ApJ, 428, L29
\bibitem[{Metzger et al.}(2008)]{Metzger08} Metzger, B. D., Piro, A. L. \& Quataert, E. 2008, MNRAS, 390, 781
\bibitem[{Metzger et al.}(2011)]{Metzger11} Metzger, B. D., Giannios,D.,Thompson,T. A.,Bucciantini, N. \& Quataert, E. 2011, MNRAS, 413, 2031
\bibitem[{Moderski et al.}(1997)]{MSL97} Moderski R., Sikora M., Lasota J. P. 1997, in Ostrowski M., Sikora M., Madejski, G., Belgelman M., eds, Proc. International Conf., Relativistic Jets in AGNs. Krakow, p. 110
\bibitem[{Mu et al.}(2016)]{Mu16} Mu, H. J., Gu, W. M., Hou, S. J. et al. 2016, ApJ, 832, 161
\bibitem[{Narayan \& Yi}(1994)]{NY94} Narayan R., Yi I., 1994, ApJ, 428, L13
\bibitem[{Narayan et al.}(2001)]{NPK01} Narayan, R., Piran, T., {\&} Kumar, P. 2001, ApJ, 557, 949 (NPK01)
\bibitem[{Novikov \& Thorne}(1973)]{NT73} Novikov, I. D., \& Thorne, K. S., 1973, in Black Holes, ed. C. DewittMorette \& B. S. DeWitt (New York: Gordon \& Breach), 345
\bibitem[{Paczy$\acute{n}$ski}(1991)]{Paczynksi91} Paczy$\acute{n}$ski, B. 1991, Acta Astron., 41, 157
\bibitem[{Paczy$\acute{n}$ski}(1998)]{Paczynksi98} Paczy$\acute{n}$ski, B. 1998, ApJ, 494, L45
\bibitem[{Page \& Thorne}(1974)]{PT74} Page, D. N., \& Thorne, K. S. 1974, ApJ, 191, 499
\bibitem[{Proga \& Zhang}(2006)]{Proga06} Proga, D., \& Zhang, B. 2006, MNRAS, 370, L61
\bibitem[{Popham \& Narayan}(1995)]{PN95} Popham, R., \& Narayan, R. 1995, ApJ, 442, 337
\bibitem[{Popham et al.}(1999)]{PWF99} Popham, R., Woosley, S. E., {\&} Fryer, C. 1999, ApJ, 518, 356 (PWF99)
\bibitem[{Pudritz \& Fahlman}(1982)]{PF82} Pudritz, R. E., \& Fahlman, G. G. 1982, MNRAS, 198, 689
\bibitem[{Reynoso et al.}(2006)]{Reynoso06} Reynoso, M. M., Romero, G. E., {\&} Sampayo, O. A. 2006, A{\&}A, 454, 11
\bibitem[{Riffert \& Herold}(1995)]{Riffert95} Riffert, H., {\&} Herold, H. 1995, ApJ, 450, 508
\bibitem[{Rosswog}(2003)]{Rosswog03} Rosswog, S., Ramirez-Ruiz, E., \& Davies, M. 2003, MNRAS, 345, 1077
\bibitem[{Rosswog}(2007)]{Rosswog07} Rosswog, S. 2007, MNRAS, 376, L48
\bibitem[{Sharkura \& Sunyaev}(1973)]{SS73} Sharkura, N. I., \& Sunyaev, R. A. 1973, A\& A, 24, 337
\bibitem[{Shibata et al.}(2007)]{Shibata07} Shibata, M., Sekiguchi, Y., \& Takahashi, R. 2007, Prog.Theor.Phys., 118, 257
\bibitem[{Tchekhovskoy et al.}(2010)]{Tchekhovskoy10} Tchekhovskoy, A., Narayan, R., \& McKinney, J. C. 2010, ApJ, 711, 50
\bibitem[{Tchekhovskoy et al.}(2011)]{TNM11} Tchekhovskoy, A., Narayan, R., \& McKinney, J. C. 2011, MNRAS Lett., 418, L79
\bibitem[{Tchekhovskoy \& McKinney}(2012)]{TM12} Tchekhovskoy, A., \& McKinney, J. C. 2012, MNRAS Lett., 423, L55
\bibitem[{Tchekhovskoy \& Giannios}(2015)]{TG15} Tchekhovskoy A., Giannios D., 2015, MNRAS, 447, 327
\bibitem[{Thorne et al.}(1986)]{Thorne86} Thorne, K. S., Price, R. H., Macdonald D. A., 1986, Black Holes: The Membrane Paradigm. Yale Univ. Press, New Haven
\bibitem[Troja et al. (2007)]{T07} Troja, E., Cusumano, G., O'Brien, P.~T., et al., 2007, ApJ, 665, 599
\bibitem[Wang et al.(2002)]{Wang02} Wang, D.~X., Xiao, K., \& Lei, W.~H.\ 2002, \mnras, 335, 655
\bibitem[{Wang et al.}(2006)]{Wang06} Wang, D.X., Lei, W.H., \& Ye, Y.C. 2006, ApJ, 643, 1047
\bibitem[{Wang et al.}(2014)]{Wang14} Wang, J.~Z., Lei, W.~H., Wang, D.~X., et al. 2014, ApJ, 788, 32
\bibitem[{Woosley}(1993)]{Woosley93}Woosley, S. E. 1993, ApJ, 405, 273
\bibitem[Wu et al.(2013)]{Wu13} Wu, X.~F., Hou, S.~J., \& Lei, W.~H.\ 2013, \apjl, 767, L36
\bibitem[{Wu et al.}(2016)]{Wu16} Wu, Q., Zhang, B., Lei, W. H., et al. 2016, MNRAS, 455, 1
\bibitem[{Xie et al.}(2016)]{XLW16} Xie, W., Lei, W.~H., \& Wang, D.~X. 2016, ApJ, 833, 129
\bibitem[{Xie et al.}(2017)]{XLW17} Xie, W., Lei, W.~H., \& Wang, D.~X. 2017, ApJ, 838, 143
\bibitem[{Xue2013}]{Xue2013} Xue, L., Liu, T., Gu, W. M. \& Lu, J. F. 2013, ApJs, 207, 23
\bibitem[{Yi et al.}(2017)]{Yi17} Yi, S. X., Lei, W. H., Zhang, B., Dai, Z. G., Wu, X. F., \& Liang, E. W. 2107, JHEAp, 13, 1
\bibitem[Yuan \& Zhang (2012)]{YZ12} Yuan, F., \& Zhang, B. 2012, ApJ, 757, 56
\bibitem[{Zalamea \& Beloborodov}(2011)]{ZB11} Zalamea, I., \& Beloborodov A. M. 2011, MNRAS, 410, 2302
\bibitem[{Zhang \& Peer}(2009)]{Zhang09} Zhang, B., \& Pe'er, A. 2009, ApJ, 700, L65
\bibitem[{Zhang et al.}(2006)]{Zhang06} Zhang, B., Fan, Y. Z., Dyks, J., et al. 2006, ApJ, 642, 354
\bibitem[{Zhang}(2007)]{Zhang07} Zhang, B. Chin. J. Astron. Astrophys. 2007, 7, 1
\bibitem[{Zhang \& Yan}(2011)]{ZY11} Zhang, B., \& Yan, H. R. 2011, ApJ, 726, 90
\bibitem[{Zhang et al.}(2008)]{Zhang08} Zhang, W., Woosley, S.~E., \& Heger, A.\ 2008, \apj, 679, 639
\bibitem[{Zhang et al.}(2017)]{ZBB17} Zhang, B. B., Zhang, B., Castro-Tirado, A. J., et al. 2017, Nature Astronomy, in press (arXiv:1612.03089)

\end{thebibliography}
\end{document}